\documentclass[useAMS,usenatbib]{mn2e}
\pdfoutput=1
\usepackage{graphicx}
\usepackage{subfigure}
\usepackage[english]{babel}
\usepackage{amssymb}
\usepackage{xspace}
\usepackage{amsmath}
\usepackage{hyperref}

\newcommand\ionm[2]{#1$\,${\small\rmfamily{#2}}}%
\newcommand{\hi}{\ionm{H}{I}\xspace}
\def\hii{\ifmmode {\mbox H{\scshape ii}}\else H{\scshape ii}\fi\xspace}
\def\h2{\ifmmode {\mbox H$_2$}\else H$_2$\fi\xspace}
\def\frach2star{$f_{\rm H2*}$\xspace}
\def\fracHIstar{$f_{\rm HI*}$\xspace}
\def\barh2frac{$\frac{\rm{M}_{\rm H2}}{\rm{M}_{\rm gas} + \rm{M}_*}$}
\def\cosmos{\ifmmode {\mbox {\it COSMOS}}\else {\it COSMOS} \fi}

\title [Gas evolution in CANDELS galaxies]{The inferred evolution of the cold gas
  properties of CANDELS galaxies at $0.5<z<3.0$}
\author[G. Popping et al.]{G.~Popping$^{1,2}$\thanks{E-mail: gpopping@eso.org},
  K.~I.~Caputi$^{2}$, S.~C.~Trager$^{2}$, R.~S.~Somerville$^{3}$,
  A.~Dekel$^{4}$,\newauthor S.~A.~Kassin$^{5}$,
  D.~D.~Kocevski$^{6}$, A.~M.~Koekemoer$^{5}$, S.M. Faber$^{7}$,
  H.~C.~Ferguson$^{5}$, \newauthor A.~Galametz$^{8}$, N.~A.~Grogin$^{5}$, Y.~Guo$^{9}$, Y.~Lu$^{10}$,
  A.~van~der~Wel$^{11}$, and B.~J.~Weiner$^{12}$\\
$^{1}$European Southern Observatory, Karl-Schwarzschild-Strasse 2,
85748, Garching, Germany\\
$^{2}$Kapteyn Astronomical Institute, University of Groningen, Postbus 800, NL-9700 AV Groningen, the Netherlands\\
$^{3}$Department of Physics and Astronomy, Rutgers University, 136 Frelinghuysen Road, Piscataway, NJ 08854, USA\\
$^{4}$Racah Institute of Physics, The Hebrew University, Jerusalem 91904, Israel\\
$^{5}$Space Telescope Science Institute, 3700 San Martin Drive, Baltimore, MD 21218, USA\\
$^{6}$Department of Physics and Astronomy, University of Kentucky, Lexington, KY 40506, USA\\
$^{7}$UCO/Lick Observatory, Department of Astronomy and Astrophysics, University of California, Santa Cruz, CA 95064, USA\\
$^{8}$Max-Planck-Institut f\"ur Extraterrestrische Physik, Giessenbachstrasse, D-85748 Garching, Germany\\
$^{9}$UCO/Lick Observatory, Department of Astronomy and Astrophysics, University of California, Santa Cruz, CA 95064, USA\\
$^{10}$The Carnegie Observatories, Pasadena, CA 91101, USA\\
$^{11}$Max-Planck Institut f\"ur Astronomie, K\"onigstuhl 17, D- 69117, Heidelberg, Germany\\
$^{12}$Steward Observatory, 933 N. Cherry St., University of Arizona,
Tucson, AZ 85721
}
\begin{document}

\maketitle

\begin{abstract}
We derive the total cold gas, atomic hydrogen, and molecular gas masses of approximately 24
000 galaxies covering four decades in stellar mass at redshifts $0.5 < z <
3.0$, taken from the CANDELS survey. Our inferences are based on the
inversion of a molecular hydrogen based star formation law, coupled with a prescription to separate
atomic and molecular gas. We find that: 1) there is an increasing trend
between the inferred cold gas (\hi and \h2), \hi, and \h2 mass and the
stellar mass of galaxies down to stellar masses of
  $10^{8}\,\rm{M}_\odot$ already in
place at $z = 3$; 2)
the molecular fractions of cold gas increase with increasing stellar
mass and look-back time; 3) there is hardly any evolution in the mean \hi
content of galaxies at fixed stellar mass; 4) the cold gas fraction and relative amount
  of molecular hydrogen in galaxies decrease at a relatively constant rate with time, independent of stellar mass;
5) there is a large population
of low-stellar mass galaxies dominated by atomic gas. These galaxies are
very gas rich, but only a minor fraction of their gas is molecular; 6)
the ratio between star-formation rate (SFR) and inferred total cold gas mass (\hi $+$ \h2)
of galaxies (i.e., star-formation efficiency; SFE)
increases with star-formation at fixed stellar masses. Due to its
  simplicity, the presented approach is valuable to assess the impact of selection biases on small samples of
directly-observed gas masses and to extend scaling relations down to
stellar mass ranges and redshifts that are currently difficult to probe
with direct measurements of gas content.
\end{abstract}

\begin{keywords}
galaxies: evolution - galaxies: formation - galaxies: ISM - ISM: molecules
\end{keywords}

\section{Introduction}
Observations in the local Universe have revealed that star formation (SF) is closely
linked to the density of molecular gas. SF in the Milky Way takes
place in dense, massive and cold giant molecular clouds
\citep{Solomon1987, McKee2007,Bolatto2008}. Recent works have emphasized
that there is a strong correlation between the star-formation rate (SFR) density and the
density of molecular hydrogen (\h2), while the correlation with the
density of atomic hydrogen (\hi) is weak or absent
\citep{Wong2002,Bigiel2008,Bigiel2011,Schruba2011}. These results have
stimulated the idea that not all the cold gas in a galaxy is necessarily
available for SF. A proper understanding of the evolution of the atomic and
molecular gas content of galaxies is a key ingredient that will shed light on the physics that regulates the stellar buildup of galaxies.

Current surveys of the cold gas content of galaxies at high redshift only probe the
molecular gas mass and are usually limited to massive galaxies
with high SFR \citep[e.g.,][]{Daddi2010,Tacconi2010,Geach2011,Bauermeister2013,Tacconi2013,Santini2013}. It is crucial to include the contribution from atomic
hydrogen to thoroughly understand how the total cold gas properties of galaxies regulate
the SFR. Furthermore, for a complete assessment it is necessary to study a galaxy sample that is not
biased to the most actively star-forming objects. It is hoped that
facilities such as ALMA (Atacama Large Millimeter Array), NOEMA (NOrthern Extended Millimeter Array), SKA (Square
Kilometer Array),  MeerKat (Karoo Array Telescope) and ASKAP (Australian SKA Pathfinder) will reveal the \hi and \h2 content of representative samples of
high-redshift galaxies.

Theorists have made considerable progress developing models that
track the \hi and \h2 content of galaxies
\citep{Obreschkow2009,Fu2010,Krumholz2011,Lagos2011cosmic_evol,Christensen2012,Kuhlen2012,Dave2013,Popping2013,
Somerville2015}. These
models have proven successful at reproducing the available
observational estimates of the overall \hi and \h2 properties of local
and high-redshift galaxies. Nevertheless,
observational constraints at high redshift are still very limited and
do not probe the wide parameter space covered by the models. Additional information on the gas content of galaxies will be
crucial to break the degeneracies in different physical mechanisms
that are included in models (e.g., SF and stellar feedback).

In the meantime, we can obtain indirect constraints on the gas content
of galaxies by using the empirical relation between SFR density and gas
density. This approach has been used extensively by inverting the
Schmidt-Kennicutt relation \citep[hereafter the KS-relation]{Kennicutt1998law}, which relates the
SFR surface density to the combined atomic and molecular hydrogen
surface density \citep[e.g.,][]{Erb2006,Mannucci2009,Troncoso2013}. \citet{Popping2012}
were the first to use an inverted molecular-gas-based SF law in
combination with a recipe to separate atomic from molecular hydrogen \citep{Blitz2006}
to estimate the total cold gas and molecular gas content of
galaxies. This approach was motivated by observations demonstrating
that SFR surface densities correlate almost linearly with
molecular gas surface density (even in the low gas surface density regime), whereas the KS-relation breaks down at low
`total gas' surface densities \citep{Bigiel2008, Schruba2011}. \citet{Popping2012}
showed that, when inferring gas masses, a molecular-gas-based SF law in combination with a
prescription to separate the atomic and molecular hydrogen content of
galaxies is better suited to reproduce directly-observed gas masses
and gas surface densities from a sample of galaxies taken from
\citet{Leroy2008} than the total-gas KS-relation. 

\citet{Popping2012} confirmed previously observed trends between
galaxy gas fraction and molecular gas fraction with stellar mass
\citep[e.g.][]{Tacconi2010,Saintonge2011}, through a detailed study of the inferred gas content of galaxies in
  COSMOS \citep[the Cosmic Evolution Survey][]{Scoville2007} at $0.5 <
z < 2.0$.  This initial study
suggested that massive
galaxies have lower gas fractions at higher redshift than less-massive
objects and have lower fractions of their gas in molecular
form.

In this paper, we apply the method developed in \citet{Popping2012}
and updated in \citet{Popping2015} to
a galaxy sample drawn from the CANDELS survey \citep[Cosmic Assembly Near-infrared Deep
Extragalactic Legacy Survey;][]{Grogin2011,Koekemoer2011} at redshifts $0.5 < z < 3$. We
focus on the cold gas (\hi $+$ \h2), \hi, and \h2 properties of galaxies over cosmic
time and how they are related to the SF and other global properties of
galaxies. The CANDELS survey is deeper than the COSMOS survey, which allows us to study
much fainter objects. As such, we can probe the gas properties of the bulk of
star-forming galaxies between $z=0.5$ and $z=3$. This cosmic epoch marks the peak in star-formation activity of our
Universe \citep{Hopkins2006,Madau2014}, when the bulk of mass in
today's massive galaxies was formed. CANDELS provides exquisite imaging covering a wide range of wavelengths to derive stellar
masses and SFR and provides reliable morphological information at these
redshifts. The
  methodology enables us to infer the gas, \hi, and \h2 masses for a
  large number of galaxies covering a wide range in stellar masses,
  SFRs, sizes, and redshift. This makes the applied method very
  helpful in assessing the impact of selection biases on much smaller
  samples for which direct gas measurements have been obtained. Furthermore, it can extend scaling relations to a stellar mass and
  redshift range difficult to reach through direct observations of gas masses. The inferred gas masses presented in this work have a great
predictive power for future \hi surveys such as LADUMA
\citep{Holwerda2012} with instruments like MeerKat, ASKAP, and the SKA
and future surveys of the molecular hydrogen content of galaxies
through CO or sub-mm continuum imaging with for example ALMA and NOEMA.

This paper is organised as follows. In Section \ref{sec:modeldata} we
summarise our method to indirectly measure the cold gas and \h2 content
of galaxies and we present the galaxy sample selection from CANDELS. In
Section \ref{sec:results} we present our results. The assumptions that
were made and the applicability of our model to the CANDELS galaxy
sample are discussed in Section \ref{sec:assumptions}. We discuss our
results in Section \ref{sec:discussion}. We summarise our findings in Section
\ref{sec:conclusion}. Throughout the paper, we assume a $\Lambda$ Cold
Dark Matter ($\Lambda$CDM) cosmology with $\rm{H}_0 =
70\,\mathrm{km\,s^{-1}\,Mpc^{-1}}$, $\Omega_{\rm matter} = 0.28$ and
$\Omega_\Lambda= 0.72$ \citep{Komatsu2009}. We assume a universal
Chabrier stellar initial mass function \citep[IMF:][]{Chabrier2003}
and where necessary convert observational quantities used to a
Chabrier IMF. All reported cold, \hi, and molecular gas masses include a correction of 1.36 to account for helium.

\section{Model \& Data}
\label{sec:modeldata}
In this section we describe our method to indirectly estimate the cold
gas (\hi $+$\h2), \hi, 
and \h2 content of galaxies and the observational data to which we
apply our method.

\subsection{Obtaining indirect gas measures}
\label{sec:model}
We infer the cold gas (\hi $+$ \h2), \hi and \h2 content of galaxies using a combination of an
empirical molecular SF law \citep[based on][]{Bigiel2008} and a
prescription to calculate the \h2 fraction of cold gas
\citep{Blitz2006}. To infer the \hi and \h2 masses of a galaxy we only use the galaxy stellar mass, SFR, and
size as input parameters. We pick a gas mass, distribute the gas as explained in the
next paragraph, and then calculate a SFR for that gas mass following
the equations given below. We repeat this process while iterating
through gas masses till convergence with the observed SFRs is
reached. When convergence is reached we separate the cold gas mass into
an atomic and molecular component using equation \ref{eq:blitz2006}.

We assume the galaxy stellar mass to be distributed following a S\'ersic
  profile $\Sigma_*(r) \propto \exp[-(r/r_*)^{1/n}]$, where $r_*$ is the
scale radius of the stellar disc for a S\'ersic profile and $n$ the S\'ersic index of the
galaxy. We also assume that the gas in galaxies is distributed following the same S\'ersic profile, with a scale radius
\begin{equation}
r_{\mathrm{gas}} = \chi_{\mathrm{gas}} \,r_*,
\end{equation}
where $\chi_{\mathrm{gas}}$ is the scale radius of the gas disc relative
to the stellar disc. We take $\chi_{\mathrm{gas}}=1.7$ based on a fit
to the galaxy disc profiles presented in \citet{Leroy2008}.

We use a slightly adapted version of the star formation law deduced by
\citet{Bigiel2008} to allow for higher star formation efficiencies in
high gas surface density regions. This is based on the results of
\citet{Daddi2010} and \citet{Genzel2010}, who found the star-formation
at high surface densities to follow the KS relation (a power-law slope of
1.4 versus 1.0 for Bigiel et al. 2008). The resulting equation is given by
\begin{equation}
\label{eq:bigiel}
\Sigma_{\mathrm{SFR}} = \frac{A_{\mathrm{SF}}}{10 M_\odot\,\mathrm{pc}^{-2}}\,\left(1 + \frac{\Sigma_{\mathrm{H2}}}{\Sigma_{\mathrm{crit}}}\right)^{N_{\mathrm{SF}}}\,f_{\mathrm{H}_2}\,\Sigma_{\mathrm{gas}}
\end{equation}
where $\Sigma_{\mathrm{SFR}}$ and $\Sigma_{\mathrm{gas}}$ are the star
formation and cold gas surface densities in
$M_\odot\,\mathrm{yr}^{-1}\,\mathrm{kpc}^{-2}$ and $M_\odot\,\mathrm{pc}^{-2}$,
respectively; $A_{\mathrm{SF}}$ is the normalization of the power law in
$M_\odot\,\mathrm{yr}^{-1}\,\mathrm{kpc}^{-2}$;
$\Sigma_{\mathrm{crit}}$ is a critical surface density; $N_{\mathrm{SF}}$ is an index which sets the
efficiency; and
$f_{\mathrm{H}_2}=\Sigma_{\mathrm{H}_2}/(\Sigma_{\mathrm{HI}}+\Sigma_{\mathrm{H}_2})$
is the molecular gas fraction. Following \citet{Popping2012} we use
$N_{\mathrm{SF}} = 0.5$ and we take $\Sigma_{\mathrm{crit}} = 70 \,M_\odot\,\rm{pc}^{-1}$.

We use a pressure-regulated recipe to determine the
molecular fraction of the cold gas, based on the work by
\citet{Blitz2006}. They found a power-law relation between the
mid-plane pressure acting on a galaxy disc and the ratio between
molecular and atomic hydrogen, i.e.,
\begin{equation}
R_{\mathrm{H}_2} = \left(\frac{\Sigma_{\mathrm{H}_2}}{\Sigma_{\mathrm{HI}}}\right) = \left(\frac{P_m}{P_0}\right)^\alpha
\label{eq:blitz2006}
\end{equation}
where $P_0$ is the external pressure in the interstellar medium where the
molecular fraction is unity; $\alpha$ is the power-law
index; and $P_m$ is the mid-plane pressure acting on the
galaxy disc. We adopted $P_0
\,=3.25\times10^{-13} \mathrm{erg\,cm^{-3}}$ and $\alpha \,=0.8$
from \citet{Leroy2008}. The mid-plane pressure can be described by \citep{Elmegreen1989}
\begin{equation}
P_m(r) = \frac{\pi}{2}\,G\,\Sigma_{\mathrm{gas}}(r)\left[\Sigma_{\mathrm{gas}}(r) + f_{\sigma}(r)\Sigma_*(r)\right]
\label{eq:pressure}
\end{equation}
where G is the gravitational constant, $r$ is the radius from the
galaxy centre, and $f_\sigma(r)$ is the ratio between
$\sigma_{\mathrm{gas}}(r)$ and $\sigma_*(r)$, the gas and stellar
vertical velocity dispersion, respectively.  Following
\citet{Fu2010}, we adopt $f_{\sigma}(r) = 0.1
\sqrt{\Sigma_{*,0}/\Sigma_*}$, where $\Sigma_{*, 0} \equiv m_*/(2 \pi
r_*^2)$, based on empirical scalings for nearby disc galaxies. 
Putting this together, we have an expression for the star formation
surface density in Equation (\ref{eq:bigiel}) depending on the cold gas surface density and stellar mass
surface density. We integrate the SFRs in the individual annuli to
obtain a total SFR which can be compared to the observed SFR.

We calibrated our method using direct observations of the \hi and/or
\h2 content of galaxies in the local and high-redshift
Universe from \citet{Leroy2008}, \citet{Daddi2010}, \citet{Tacconi2010}, and
\citet{Tacconi2013}\footnote{\citet{Leroy2008},\citet{Tacconi2010}, and
  \citet{Tacconi2013} assume a CO-to-\h2 conversion factor of $X_{\rm
    CO} = 2\times10^{20}
  \mathrm{cm^{-2}/(K\,km\,s^{-1})}$. \citet{Daddi2010} assumes a
  CO-to-\h2 conversion factor of $X_{\rm
    CO} = 2.25\times10^{20}
  \mathrm{cm^{-2}/(K\,km\,s^{-1})}$}. Using $\chi^2$-minimization, we find the best agreement between predicted and observed gas
masses when adopting a value of $A_{\mathrm{SF}}=9.6 \times
10^{-3}\,M_\odot\,\mathrm{yr}^{-1}\,\mathrm{kpc}^{-2}$. This value is
close to normalizations found in \citet{Bigiel2008} and
\citet{Bigiel2011}, $8.42\times 10^{-3}$ and $4.6\times
10^{-3}\,M_\odot\,\mathrm{yr}^{-1}\,\mathrm{kpc}^{-2}$, respectively. We integrate the gas disc out to 10 times
$r_{\rm gas}$. We discuss the uncertainties on some of the
individual components of this model in Section
\ref{sec:assumptions}. The typical systematic uncertainty of our
method is 0.3 dex.  
\begin{figure*}
\includegraphics[width = 0.9\hsize]{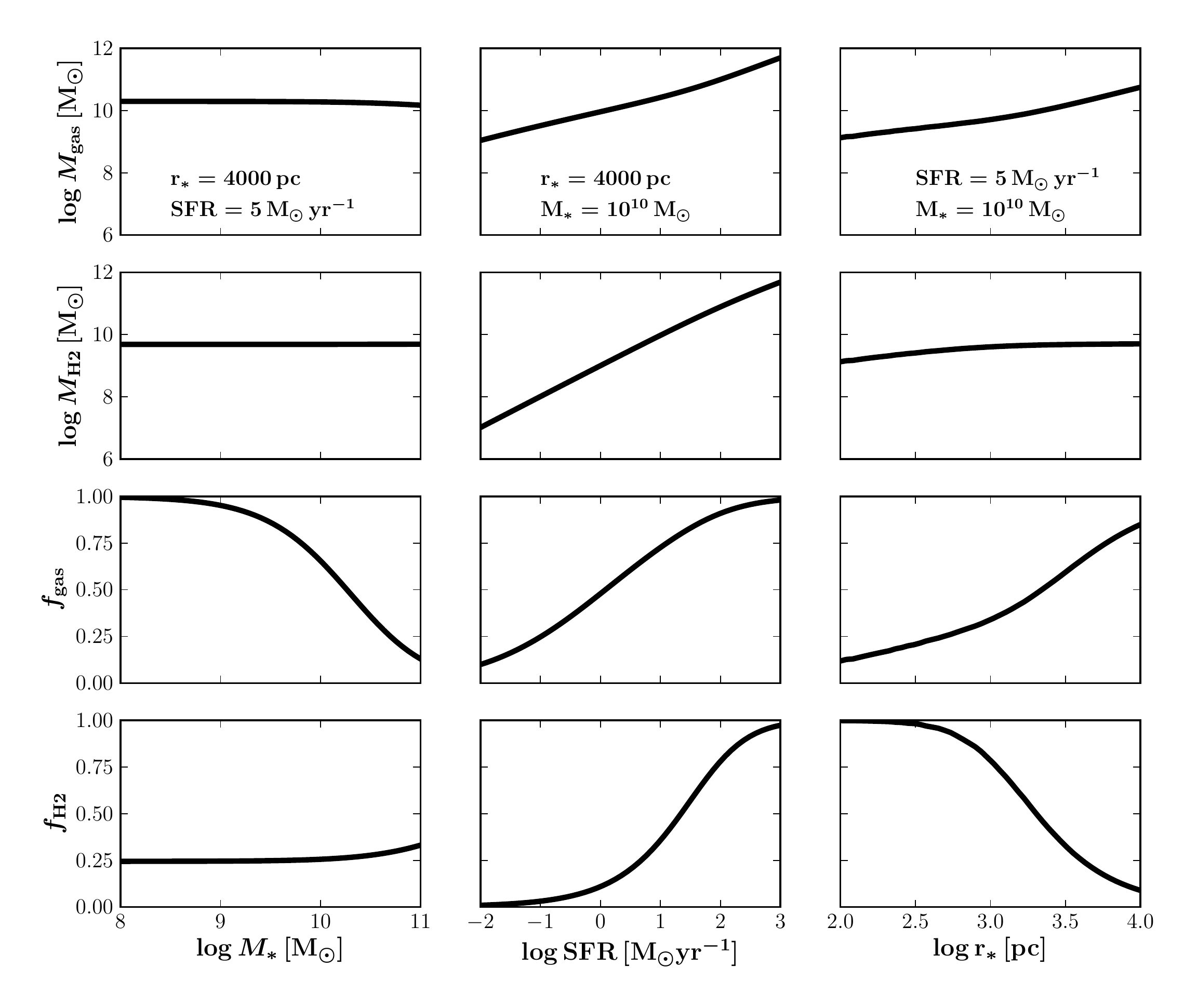}
\caption{ Schematic change in cold gas mass, \h2 mass, cold gas fraction ($f_{\rm
    gas} \equiv M_{\rm gas}/(M_{\rm gas} + M_*)$) and molecular hydrogen
  fractions ($f_{\rm H2}\equiv M_{\rm H2}/M_{\rm gas}$) produced by our model when varying the
  stellar mass (left column), SFR (middle column) and scale radius
  (right column) of a galaxy. For each column the other two of three
  input galaxy properties of our model (stellar mass, SFR, and scale
  radius) are fixed. The fixed values are quoted in the upper panels. \label{fig:model_behavior}}
\end{figure*}

A schematic picture of the dependencies of our model is presented in Figure
\ref{fig:model_behavior}. In this figure we explore the effects of
changing one of the three input properties of a galaxy (SFR, size, and stellar
mass) on the estimated cold gas mass and \h2 mass of a galaxy and the cold gas
fraction ($f_{\rm gas}$) and molecular hydrogen fraction of the cold gas ($f_{\rm H2}$). We keep two of
the parameters fixed and vary the third. Changing the stellar mass
of the galaxy has a negligible effect on the inferred cold gas and
\h2 mass. It is only for the most massive galaxies ($M_* \approx
10^{11}\,\rm{M}_\odot$) that, due to the increased pressure from stars,
the inferred cold gas mass decreases and $f_{\rm H2}$
increases. Because the inferred cold gas mass remains relatively
constant, the cold gas fraction $f_{\rm gas}$ rapidly decreases
with increasing stellar mass. 

We find a strong increase in cold gas
mass, \h2 mass, $f_{\rm gas}$, and $f_{\rm H2}$ when increasing the SFR. This is not surprising, as more gas is needed to
support the higher SFR. The increased gas mass enhances the cold
gas surface density and pressure acting on the gas, which leads to
an increase in $f_{\rm H2}$. 
\begin{figure}
\centering
\includegraphics[width = 0.9\hsize]{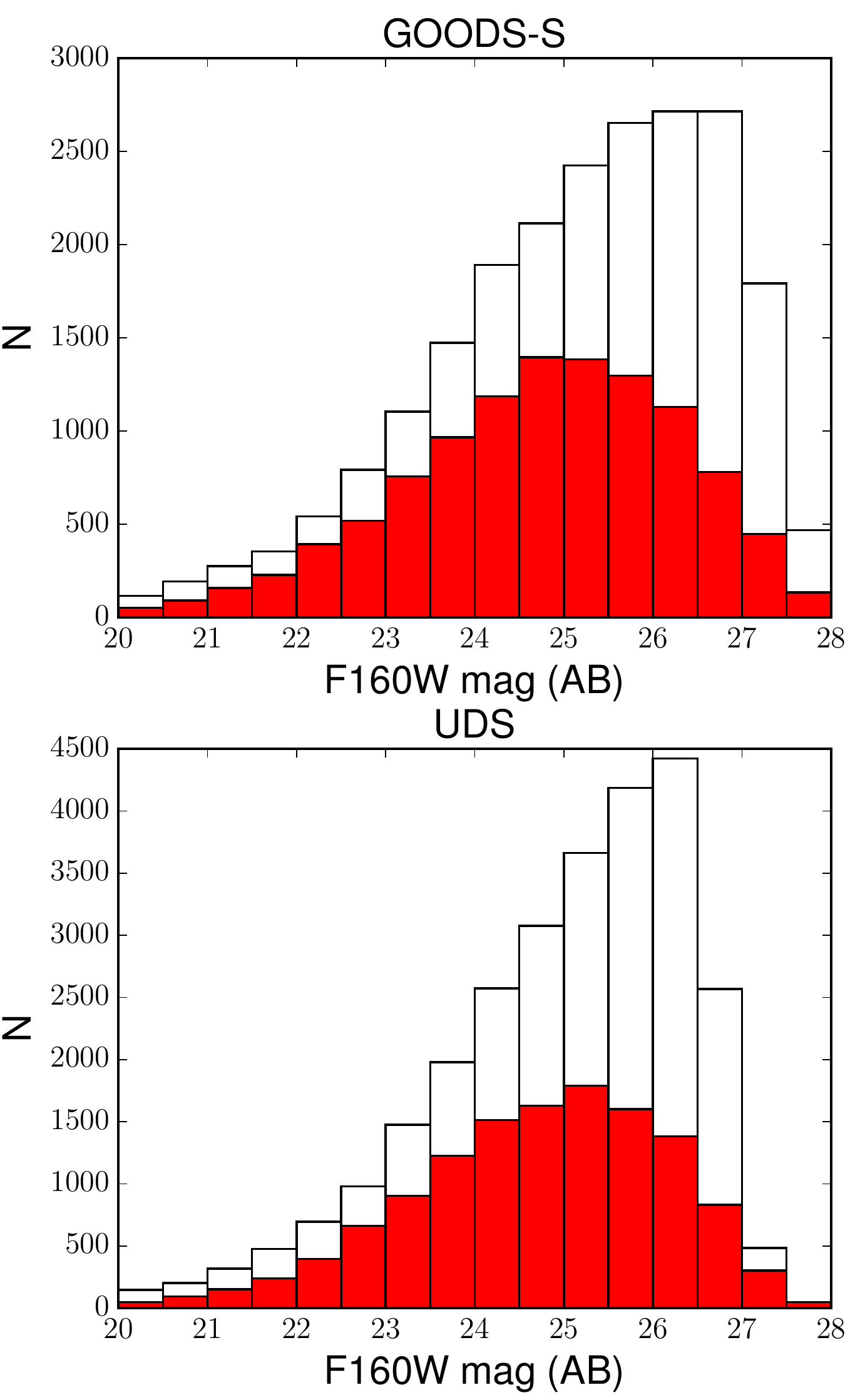}
\caption{$H$-band magnitude distributions of the original (white) and
  final samples (after the morphology cut; red) at redshifts $0.5 < z <
  3$ for the GOODS-S (top) and UDS (bottom) fields. \label{fig:data}}
\end{figure}

The inferred gas masses increase as a
function of scale radius. When increasing the scale radius, the gas
densities decrease, therefore the molecular fraction of the gas
decreases and the SFR surface density decreases as well. To ensure
that convergence with the input SFR is obtained, more gas has to be
added to the galaxy to make up for the lowered \h2 fractions. This
results in high gas fractions and low molecular hydrogen fractions.

\subsection{Data}
\label{sec:data}
We infer the galaxy gas content of two deep
samples of F160W ($H$-band) selected galaxies from the CANDELS survey at
$0.5<z<3$ \citep{Grogin2011,Koekemoer2011}. One of the samples corresponds to the Great Observatories Origins
Deep Survey South \citep[GOODS-S;][]{Giavalisco2004}, and the other to the
UKIDSS Ultra-Deep Survey (UDS) field \citep{Lawrence2007}. We used the
publicly available CANDELS photometric catalogues compiled
by \citet{Guo2013} and \citet{Galametz2013}, and the redshift
compilation by \citet{Dahlen2013}, with spectroscopic redshift updates
taken from \citet{Fadda2010} and \citet{Barro2014}, to select our
galaxies.  We selected disc galaxies based on their 
S\'ersic index and excluded galaxies that
contain an active galactic nucleus (AGN). We did not apply any
selection based on stellar mass or SFR. The exact selection criteria
and our methods to calculate stellar masses, SFRs, and sizes are
described below.

As our method requires us to know the scale radius of each galaxy,
we extracted from these catalogues all those sources with good-quality
morphological fits in the $H$ band, as determined by
\citet{vanderWel2012}. We computed the circularized, effective radius
$r_e$ of a galaxy as $r_e = a_e \sqrt{(b/a)}$, where $a_e$ is the
effective radius along the major axis and $b/a$ the ratio between the
minor and major axis sizes and converted these to the scale
length $r_*$ for a S\'ersic profile with S\'ersic index $n$. We converted all scale radii to a common rest-frame
wavelength of  5000 {\AA} following \citet{vanderWel2014}. Since
our model is designed for disc galaxies, we discarded all galaxies with a S\'ersic index
$n>2.5$. 

We excluded galaxies known to contain an AGN from
our sample, to avoid any bias in the derived star formation rates and
stellar masses.  To identify AGN, we used the 4Ms Chandra X-ray catalogue
for the GOODS-S field \citep{Xue2011}, and an infrared power-law
spectral energy distribution analysis \citep{Caputi2013} in the UDS field, as
the existing X-ray data is shallow \citep{Ueda2008}.  These AGN constitute only 1--2\%
of the CANDELS $0.5<z<3$ samples. It is possible that other galaxies
containing an AGN remained in our sample, but the AGN is likely weak
as it is not identified in the X-ray data or with the IRAC power-law criterion.

By applying these morphology and AGN cuts, we kept  around 60\% of the galaxies in the GOODS-S and UDS  CANDELS catalogues. The
distributions of the $H$-band magnitudes of the original and final samples at $0.5<z<3$ for the two
fields separately are compared in Fig. \ref{fig:data}. From this Figure, we can see
that 1) the GOODS-S sample reaches around a magnitude deeper than the UDS
sample, as the former is based on both the deep and ultra-deep CANDELS
data for the GOODS-S field; and 2) the percentage of galaxies lost
because of the good-quality morphology cut and cut in S\'ersic index
is not constant with $H$-band magnitude. This cut still contains
$>50\%$ of the original galaxies for $H<26$, but the sampling becomes progressively more sparse at
fainter magnitudes. The depth of the GOODS-S sample reflects
itself in a relatively higher number of low-stellar mass
galaxies. Nevertheless, our sample is likely to miss extended
galaxies with low surface brightness. This becomes especially
important for galaxies with low stellar masses at $z > 2$. For our
model this means that we are biased against low surface
brightness galaxies \citep{Guo2013}.

We computed stellar masses using a multi-wavelength SED $\chi^2$ fitting to the
CANDELS photometry (from the $U$ to the $4.5\,\mu \rm{m}$ band), applying the
\citet{bruzual2007} templates fixed at the redshifts of
the sample objects. We used a single stellar population and five exponentially-declining star formation
histories (with $\tau = 0.1,\,0.5,\,1.0,\,2.0,\,\rm{and}\,5.0$
Gyr). For each star formation history we considered 24 possible
templates, corresponding to ages between 0.05 and 5.0 Gyr. The
\citet{bruzual2007} templates have a larger contribution from
thermally pulsing asymptotic giant branch (TP-AGB) stars than the
\citet{bruzual2003} templates. Stellar masses calculated based on
\citet{bruzual2007} are on average 0.1 dex less massive than stellar
masses calculated based on the \citet{bruzual2003} models. Internal dust
extinction has been taken into account by convolving all Bruzual and
Charlot templates with the \citet{Calzetti2000} reddening law, with $A_{\rm v}$
values ranging from 0.0 to 3.0 (with a step of 0.1). 

SFRs are based on rest-frame UV fluxes corrected for extinction E(B-V) for each
individual galaxy using the \citet{Calzetti2000} reddening law. UV fluxes were converted into SFRs following
\citet{Kennicutt1998SFR} for a Chabrier IMF. When IR photometry was
available (for just a few percent of our sample) we
calculated SFRs based on a combination of non-extinction corrected UV and IR
photometry. We derived rest-frame $8\mu$m luminosities from
the $24\mu$m fluxes, and converted the $8\mu$m luminosities into total
IR luminosities following \citet{Bavouzet2008}. SFRs were calculated
following \citet{Kennicutt1998SFR} and \citet{Bell2005}. In comparison
with \citet{Popping2012} the CANDELS survey is closer to a
mass-selected sample and (when IR photometry is available) is based on
a more-reliable tracer of the SFR. As such, the results presented in
this work supercede our previous efforts.

\begin{figure*}
\includegraphics[width = 0.9\hsize]{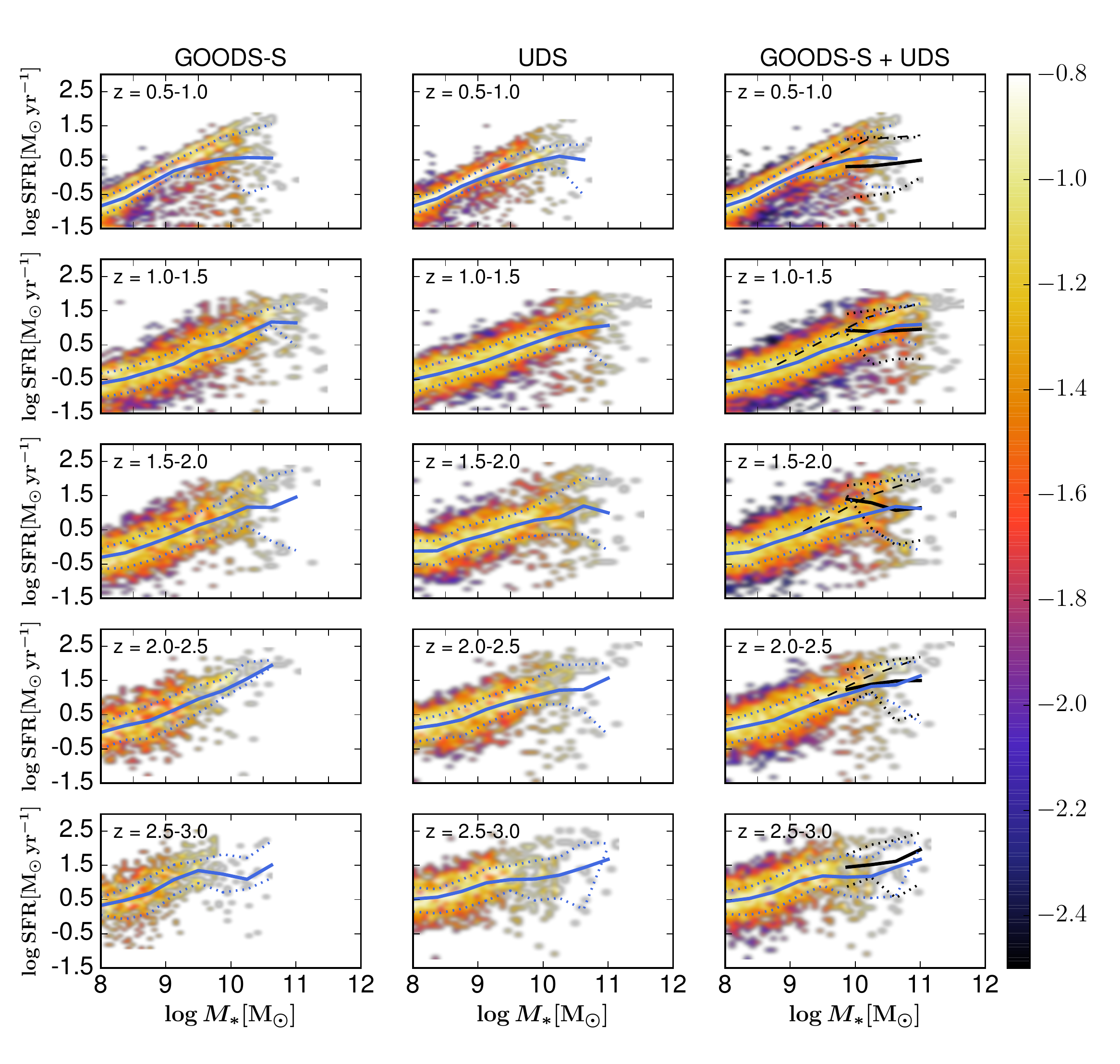}
\caption{Galaxy SFR as a function of stellar mass in different redshift bins. The shaded
  regions show the log of the conditional probability distribution
  function $P({\rm SFR}|M_*)$, which represents the probability
  distribution of SFRs for fixed stellar masses, and the blue solid and dotted lines show
  the median fit and $2\sigma$ deviation. The left column shows results for the GOODS-S
sample, the middle column for the UDS sample, and the right column for the
combined GOODS-S and UDS sample. The black
line marks the mean trend of the UV + IR based SFRs presented in
\citet{Barro2014} in the GOODS-S and UDS fields. The black dashed line
mark the double-power-law fit to the relation between stellar mass and
SFR presented in \citet{Whitaker2014}.
\label{fig:mstar_sfr}}
\end{figure*}

 \section{Results}
\label{sec:results}
In this section we present galaxy SFRs and the inferred total cold
gas, atomic hydrogen, and molecular hydrogen masses of galaxies in the CANDELS survey. We
will focus on the gas masses, gas fractions, gas properties of
galaxies and the evolution of the gas content of galaxies. 

\subsection{Galaxy SFR}
To fully appreciate the predictive power of our model it is
crucial to place the inferred gas masses in their proper context. Our cold gas and \h2
estimates depend on the SFR of a
galaxy to first order. In Figure \ref{fig:mstar_sfr} we present the SFR of our galaxy
sample as a function of stellar mass for different redshift bins. For clarity the relations are shown for the
samples taken from the GOODS south field, the UDS field, and the
combination of these fields. We
find an increasing trend in SFR with stellar mass up to
redshifts $z<3$. The median of this trend has been referred to as the
`main sequence' of star-forming galaxies
\citep[e.g.,][]{Noeske2007,Elbaz2011}. The
normalization of the trend between stellar mass and SFR decreases with
time. At fixed stellar mass, galaxies in the redshift range $2.5 < z
<3.0$ formed stars an order of magnitude more rapidly than in the
redshift range $0.5<z<1.5$. At $0.5<z<1.0$ the increasing trend of SFR
with stellar mass has a slope 
of 0.66. Other studies have found a slope ranging from 0.6 to 1 at the same redshifts
\citep{Pannella2009,Karim2011,Whitaker2012,Whitaker2014}. 

We did not apply any additional
selection criteria beyond those described in Section
\ref{sec:data}. While many other studies only select galaxies on the
`main sequence' of
star-formation, our sample also includes a fraction of
galaxies with low SFRs.

We compare our SFRs with SFRs from \citet[black line in Figure
\ref{fig:mstar_sfr}]{Barro2014} and the double power-law fit to the
stellar mass -- SFR relation presented in \citet[black dashed
line]{Whitaker2014}. We find that our mean trends for the
stellar mass -- SFR relation are in reasonable agreement with trends
found in the literature. We note that \citet{Barro2014} and
\citet{Whitaker2014} applied criteria to select actively star-forming
galaxies, attempting to exclude more quiescent objects, whereas our selection
criteria do not rule out quiescent objects.

There  is a large population of galaxies with stellar masses larger
than $\sim 10^{10}\,M_\odot$  and low SFRs
($<1.0\,\rm{M}_\odot\,\rm{yr}^{-1}$). These galaxies with low SFRs are less actively forming
stars than counterparts at fixed stellar mass and make up a group
of quiescent galaxies \citep[]{Brennan2015}. 
For these stellar masses our
  sample is significantly incomplete (completeness $< 50$ per cent)
  below SFRs of $\log(\mathrm{SFR}/M_\odot\,\mathrm{yr^{-1}})=-0.5$ at
  $z=2$. The SFR limit to observe galaxies increases at higher redshifts. This implies that at redshifts $z>2.0$ the
  contribution by galaxies with low SFRs may be larger. 

\begin{figure*}
\includegraphics[width = 0.9\hsize]{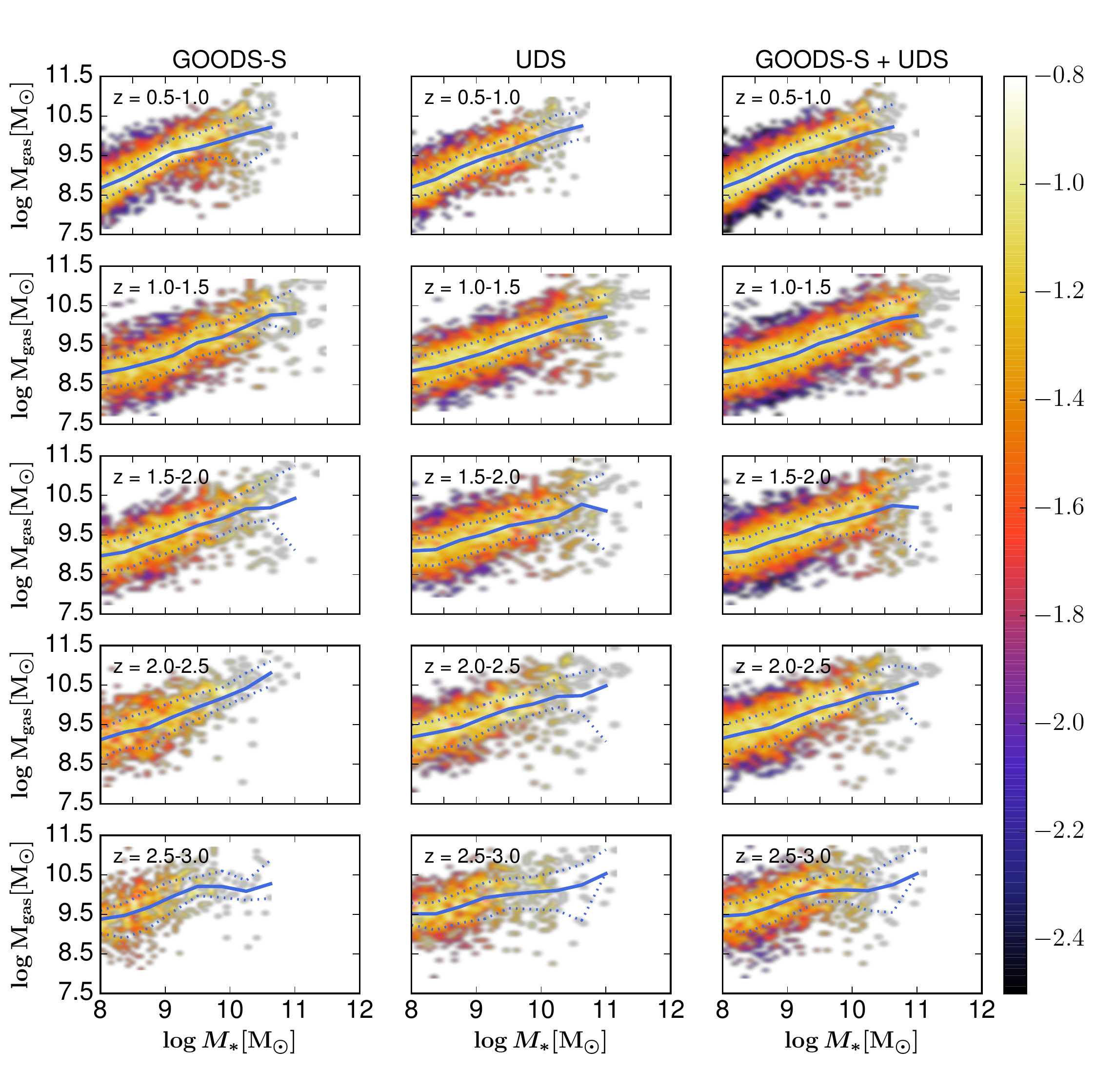}
\caption{Total cold gas mass of galaxies as a function of their
  stellar mass in different redshift bins. The shaded
  regions show the log of the conditional probability distribution
  function $P({\rm M}_{\rm{gas}}|M_*)$, and the blue solid and dotted lines show
  the median fit and one $\sigma$ deviation.  Columns are as described in the
  caption of Figure \ref{fig:mstar_sfr}.
\label{fig:mstar_mgas}}
\end{figure*}

\begin{figure*}
\includegraphics[width = 0.9\hsize]{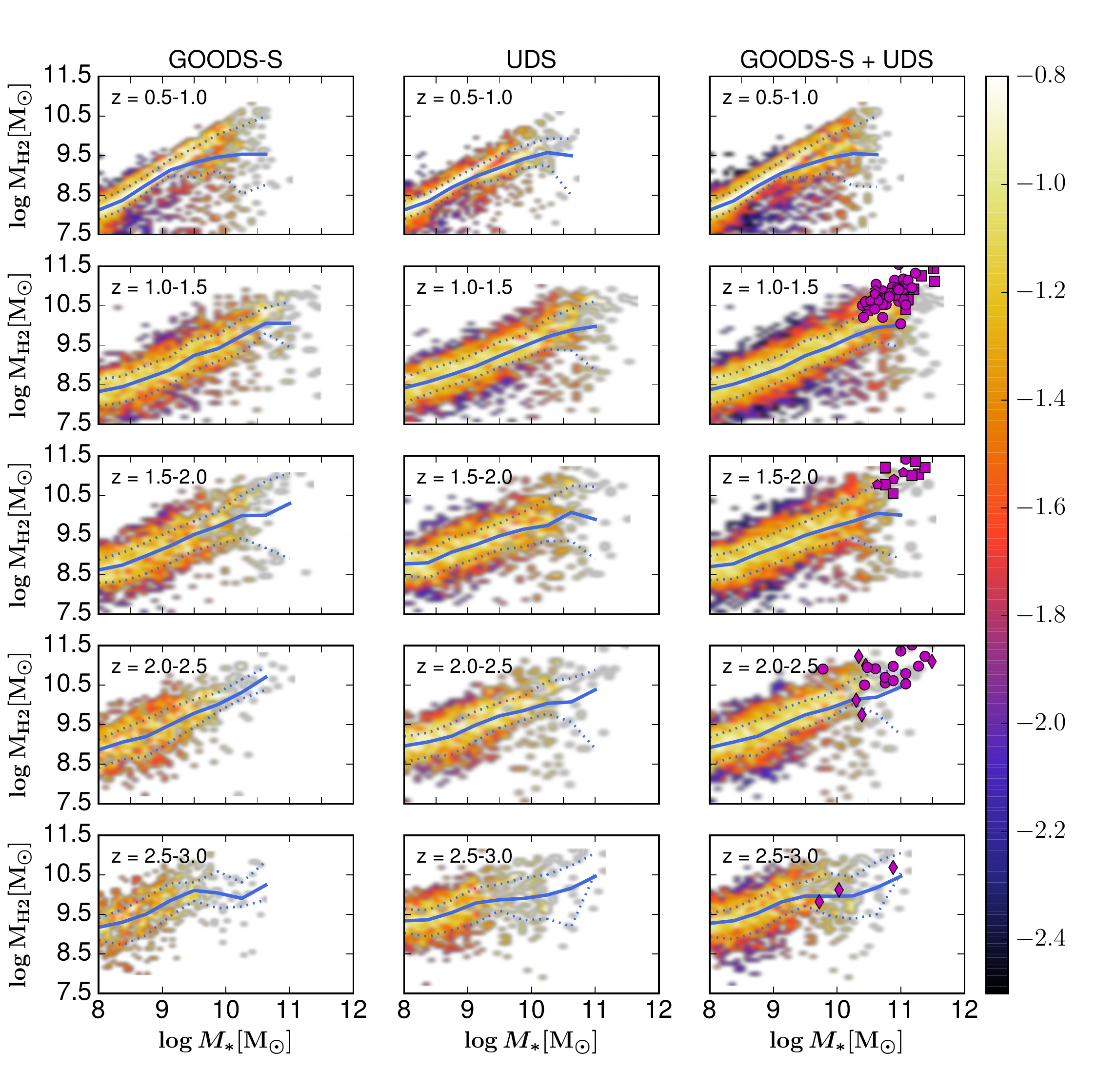}
\caption{\h2 mass of galaxies as a function of their stellar mass
  different in redshift bins. The shaded
  regions show the log of the conditional probability distribution
  function $P({\rm M}_{\rm{H2}}|M_*)$, and the blue solid and dotted lines show
  the median fit and one $\sigma$ deviation.  Columns are as described in the
  caption of Figure \ref{fig:mstar_sfr}. Purple pentagons, circles,
  squares, and diamonds are literature values from \citet{Daddi2010}, 
  \citet{Tacconi2010}, \citet{Tacconi2013}, and \citet{Saintonge2013}, respectively.
\label{fig:mstar_mh2}}
\end{figure*}

\begin{figure*}
\includegraphics[width = 0.9\hsize]{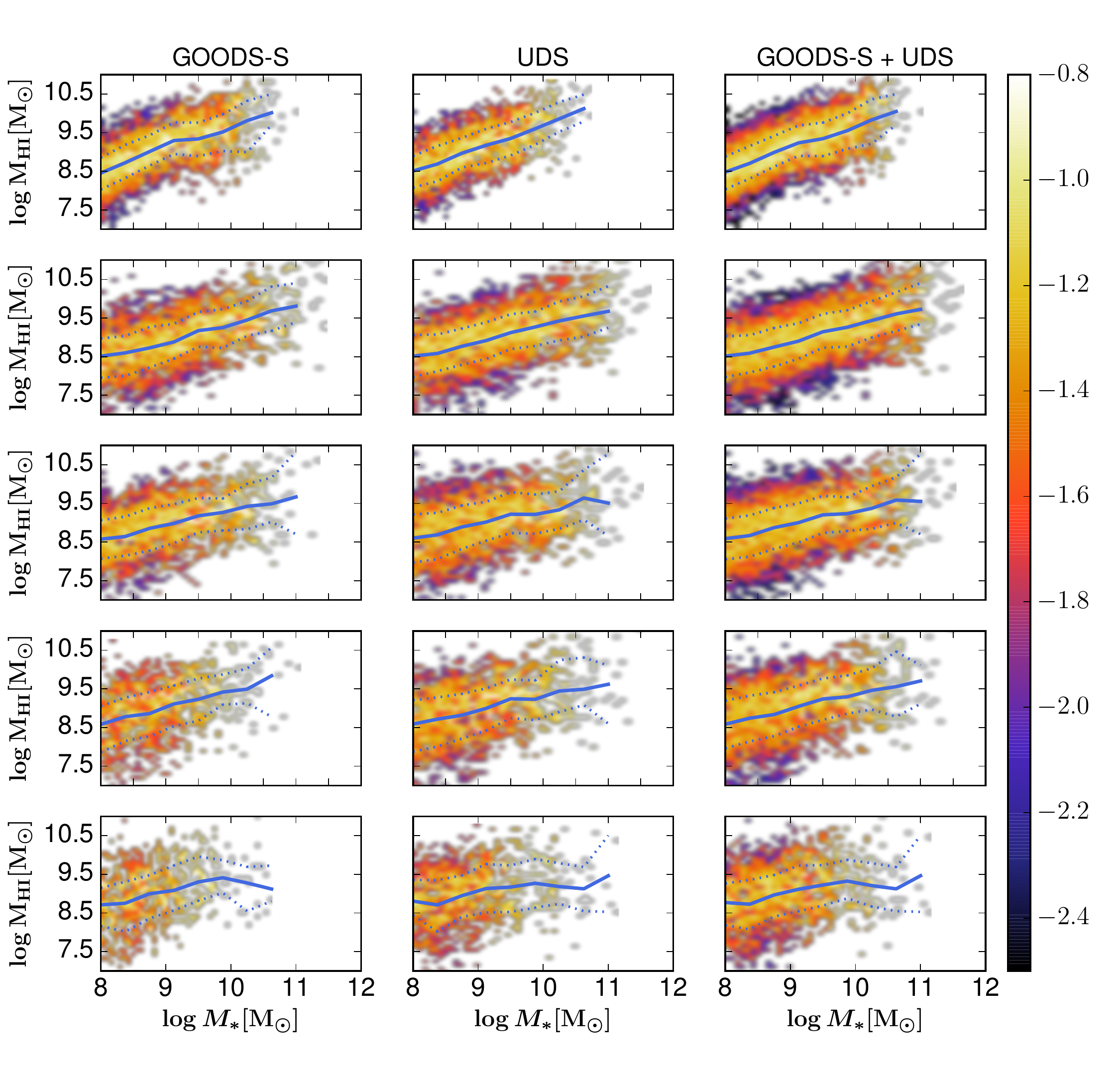}
\caption{The \hi mass of galaxies as a function of their
  stellar mass in different redshift bins. The shaded
  regions show the log of the conditional probability distribution
  function $P({\rm M}_{\rm{HI}}|M_*)$, and the blue solid and dotted lines show
  the median fit and one $\sigma$ deviation. Columns are as described in the
  caption of Figure \ref{fig:mstar_sfr}.
\label{fig:mstar_mHI}}
\end{figure*}

\subsection{Cold gas, \hi, and \h2 content}
We present the derived total cold gas masses of our
galaxy sample as a function of stellar mass in different redshift
bins in Figure \ref{fig:mstar_mgas}. We find that on average the cold
gas mass of a galaxy increases with its stellar mass at all observed
redshifts with a sub-linear slope. Despite the clear increasing trend, we do find a significant
amount of scatter. Note especially a group of gas-poor galaxies with high stellar
masses. This group of galaxies seems to have decoupled itself from the general increasing
trend of gas mass with stellar mass, indicating that some physical
process has either removed some of the cold gas or prohibits the cooling
and/or accreting of gas onto the galaxy. These galaxies are likely on
their way towards the red sequence. There is a shift in the normalization in the relation between
gas- and stellar-mass with redshift. On average, high-redshift galaxies
(especially at $z>2.0$) are more gas-rich than galaxies with similar
stellar mass at lower redshifts. At $0.5<z<1.0$ the increasing trend
of cold gas mass with stellar mass has a slope
of 0.58. 

We present the derived \h2 masses of our galaxy sample in five
redshift bins in Figure \ref{fig:mstar_mh2}. Similarly to the total cold
gas mass, we find an increase in \h2 mass with stellar mass and a
decrease in \h2 mass with decreasing redshift. We see a group of
galaxies with high stellar masses and low \h2 masses at $z<2.5$ in good
agreement with a similar group of galaxies seen in the relation between
cold gas and stellar mass (Fig. \ref{fig:mstar_mgas}). The relation between \h2 mass and stellar
mass is well defined and has a clear upper envelope over the entire range of stellar
masses probed, with only minor stochastic appearance of galaxies above the
envelope. The trend between \h2 mass and stellar mass is slightly steeper than for
the total cold gas mass. At
$0.5<z<1.0$ the increasing trend of \h2 mass with stellar mass has a slope
of 0.65. Although the relations between
stellar mass and cold and \h2 mass are very similar, there is no simple
one-to-one mapping between cold gas mass and \h2 mass.

As a comparison we have included direct measures of the \h2 mass of galaxies (through their
CO luminosity) from the literature
\citep{Tacconi2010,Saintonge2013,Tacconi2013} in Figure
\ref{fig:mstar_mh2}.\footnote{\citet{Saintonge2013} calculates
  the CO-to-\h2 conversion factor as a function of metallicity using
  the prescription presented in \citet{Genzel2012}.} The mean trend
for the inferred \h2 masses lies below the observed sample. Our sample contains 24000 galaxies covering four decades in stellar
mass. The observational data points correspond to only a small number
of galaxies in a very localized region of parameter space. Our sample
was not selected to purely consist of `main sequence'
galaxies, driving the mean trends in inferred gas mass down with
respect to a galaxy sample consisting of only `main sequence' galaxies
only.  A close look at Figure \ref{fig:mstar_mh2} shows that  most
of the directly-observed \h2 masses are in good agreement with the
most-\h2-massive (i.e., actively star-forming) galaxies at stellar masses
$M_* > 10^{10}\,\rm{M}_\odot$. \footnote{\citet{Tacconi2013} selected galaxies with a
  stellar masses $\geq 2.5\times 10^{10}\,\rm{M}_\odot$ and
  star-formation rate$ \geq
  30\,\mathrm{M_\odot yr^{-1}}$. This matches the most-actively
  star-forming galaxies in our sample.} 
We find that
the increasing relation between stellar mass and \h2 mass suggested by
the direct observations continuous at least down to stellar masses of $10^8\,\rm{M}_\odot$.

We show the \hi mass of galaxies as a function of their stellar mass
in different redshift bins in Figure \ref{fig:mstar_mHI}. Like for the
cold gas and \h2 mass, the \hi content of galaxies increases as a
function of stellar mass. The scatter around the mean trend is very wide,
much wider than trends with cold gas mass and \h2 mass. At $0.5 < z <
1.0$ the trend between \hi mass and stellar mass has a slope of 0.54.

The relations between stellar mass and cold gas, \hi, and \h2 mass tightly
follow the trend between SF and stellar mass (Figure
\ref{fig:mstar_sfr}). The well defined upper envelope of this
relation for non-starburst galaxies is in good agreement with the
upper envelope of the relation between stellar mass and \h2 mass. It
is important to realize that in our model the \h2 mass is {\it to first order} a different
representation of the SFR of a galaxy (through $A_{\rm SF}$), hence the similarity between the slopes of the relation
of SFR and \h2 mass with stellar mass (0.66 for SFR versus 0.65 for the \h2 mass at $z=0.5-1.0$). The
total cold gas mass and \hi mass are driven through a more complex combination of
stellar mass, SFR and disc size, and therefore do not necessarily
have a simple relationship with SFR. This is reflected in the difference in
the slopes of the relations between stellar mass and SFR (slope of
0.66), cold gas mass (slope of 0.58), and \hi (slope of 0.54). The inferred cold gas mass has
a slightly shallower increase with stellar mass than the inferred \h2 mass
and SFR.

\begin{figure}
\includegraphics[width = 0.9\hsize]{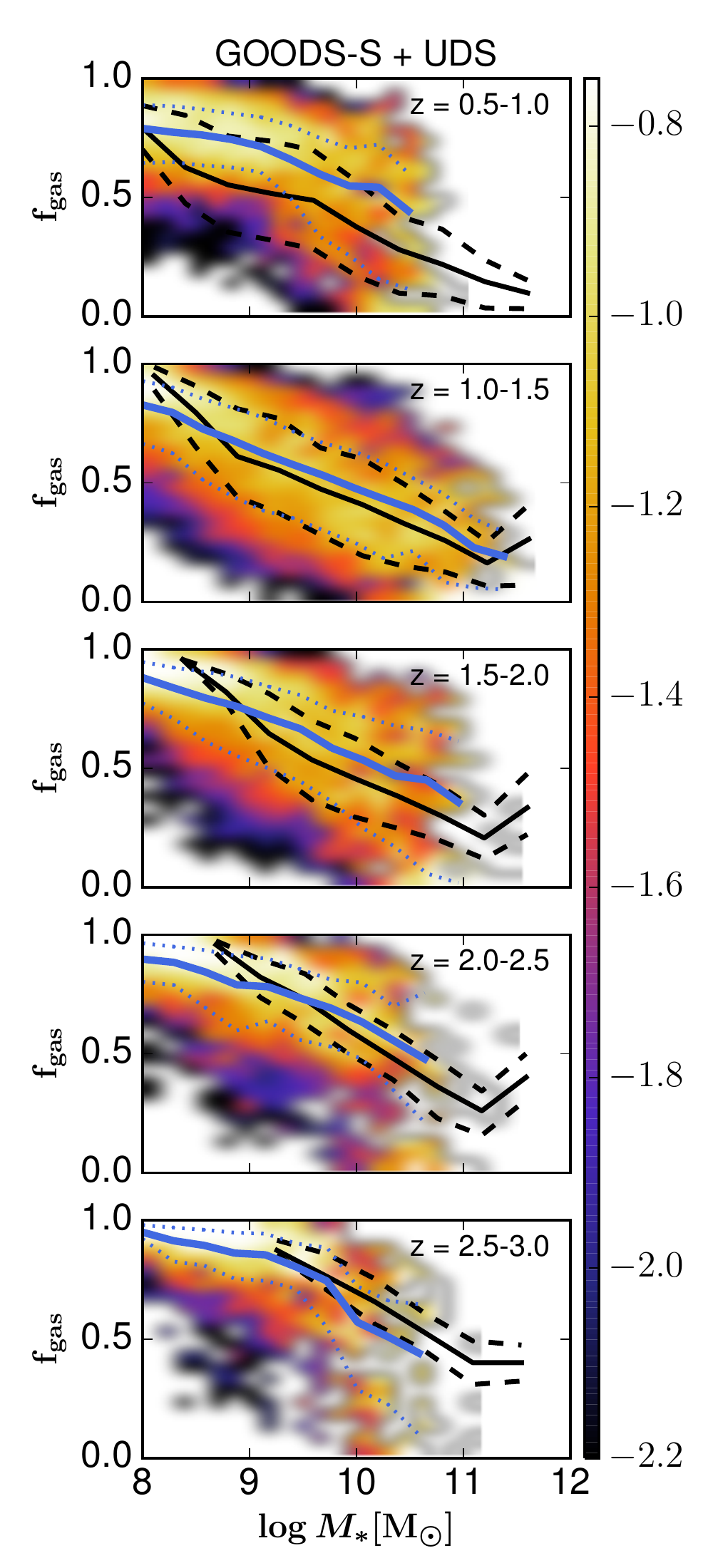}
\caption{Cold gas fraction of galaxies  ($f_{\rm{gas}} \equiv
\frac{M_{\rm{H2 + HI}}}{M_{\rm{H2 + HI}} + M_*}$) as a function of
stellar mass and redshift. The shaded
  regions show the log of the conditional probability distribution
  function $P(f_{\rm{gas}}|M_*)$, and the blue solid and dotted lines show
  the median fit and one $\sigma$ deviation.  The black solid and dashed
  lines show the mean fit and one $\sigma$ deviation to the
  predictions of \citet{Popping2013}.\label{fig:fgas}}
\end{figure}
\begin{figure}
\includegraphics[width = 0.9\hsize]{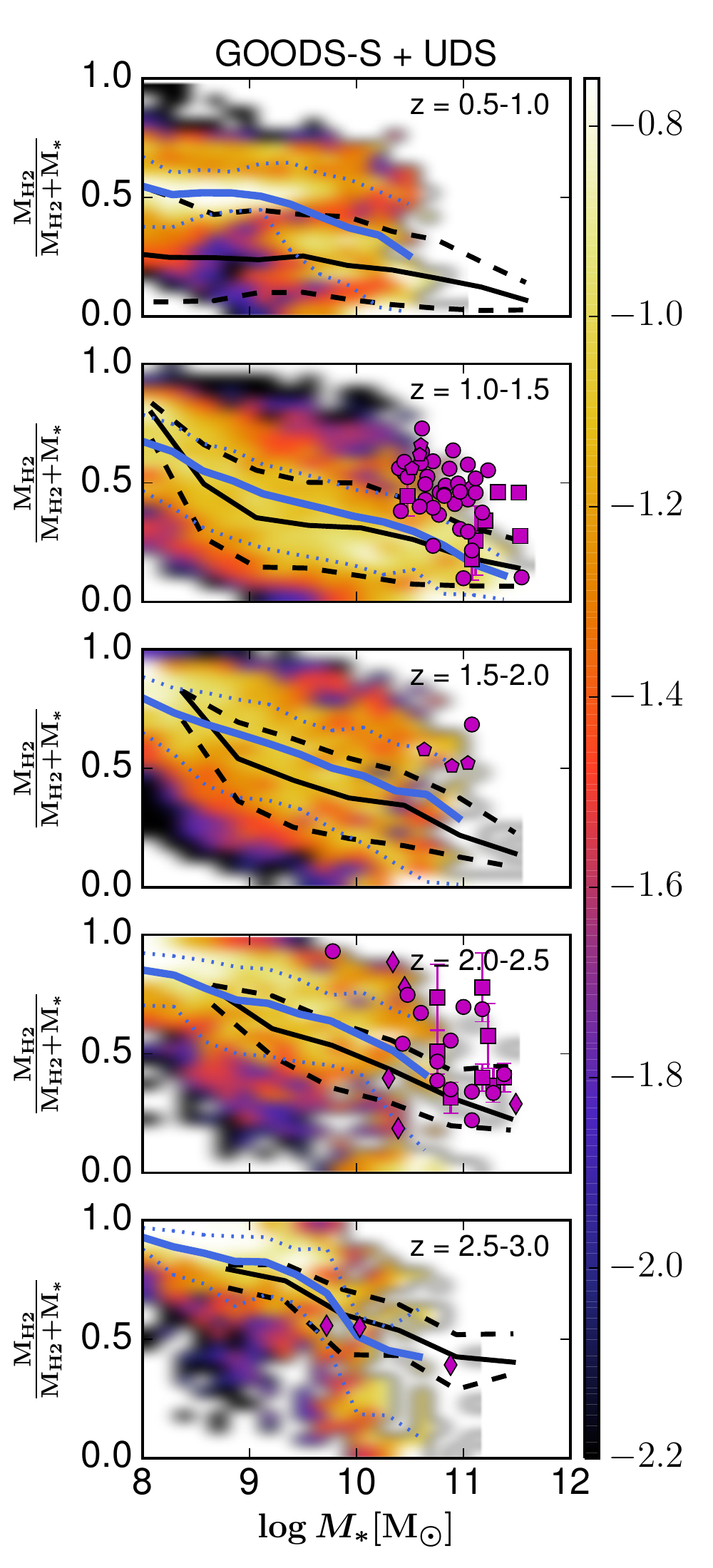}
\caption{Relative molecular content as a function
  of stellar mass for different redshift
  bins.  The shaded region shows the
  log of the conditional probability distribution function
  $P(\frac{\rm{M}_{\rm H2}}{\rm{M}_{\rm H2} + \rm{M}_*}|M_*)$, and the blue solid and dotted lines show
  the median fit and one $\sigma$ deviation. The black solid and dashed
  lines show the mean fit and  1 $\sigma$ deviation to the
  predictions of \citet{Popping2013}. Purple pentagons,
  circles, squares and diamonds are literature values from
  \citet{Daddi2010}, 
  \citet{Tacconi2010}, \citet{Tacconi2013}, and
  \citet{Saintonge2013}, respectively.
\label{fig:fh2_star}}
\end{figure}

\begin{figure}
\includegraphics[width = 0.9\hsize]{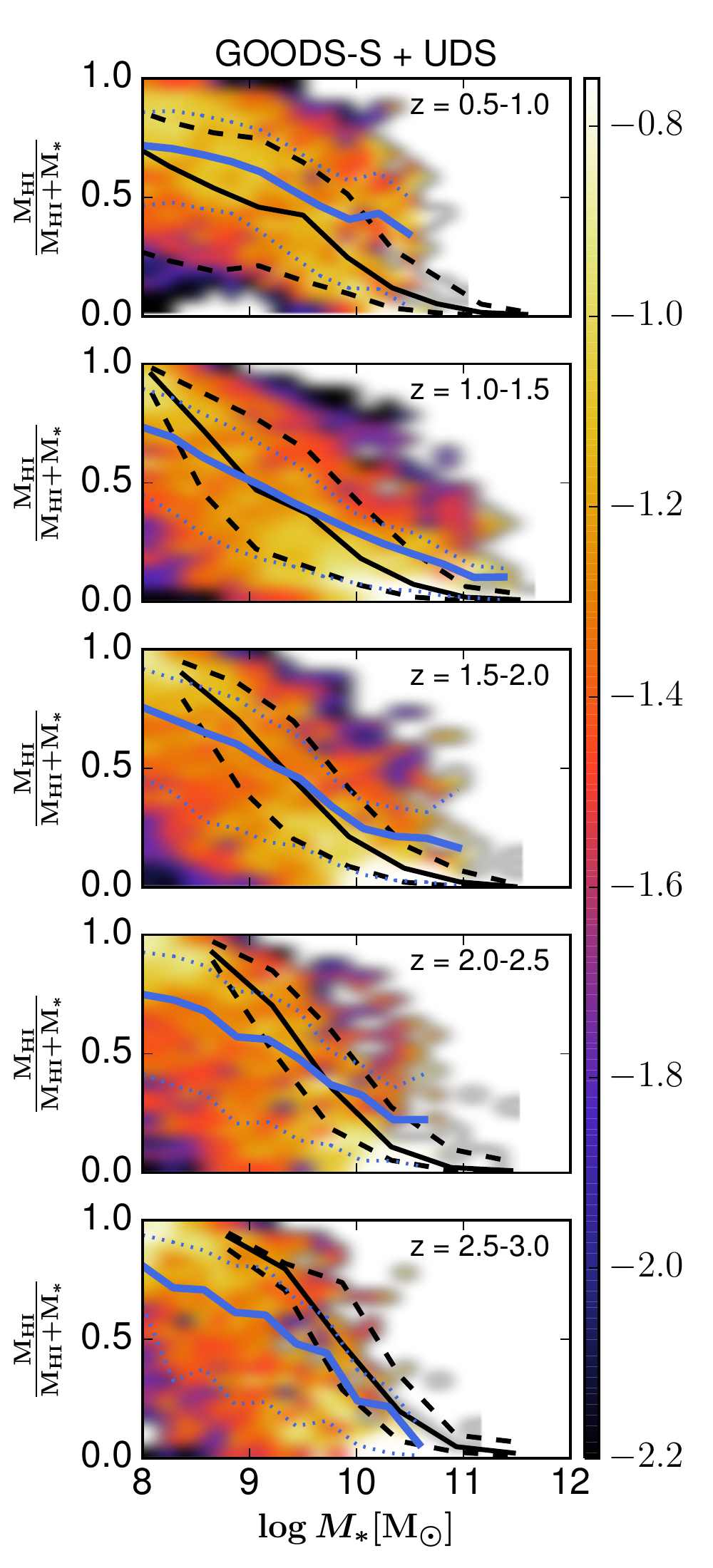}
\caption{Relative atomic hydrogen content as a function
  of stellar mass for different redshift
  bins.  The shaded region shows the
  log of the conditional probability distribution function
  $P(\frac{\rm{M}_{\rm HI}}{\rm{M}_{\rm HI} + \rm{M}_*}|M_*)$, and the blue solid and dotted lines show
  the median fit and one $\sigma$ deviation. The black solid and dashed
  lines show the mean fit and  1 $\sigma$ deviation to the
  predictions of \citet{Popping2013}. 
\label{fig:fracHIstar}}
\end{figure}

\begin{figure}
\includegraphics[width = 0.9\hsize]{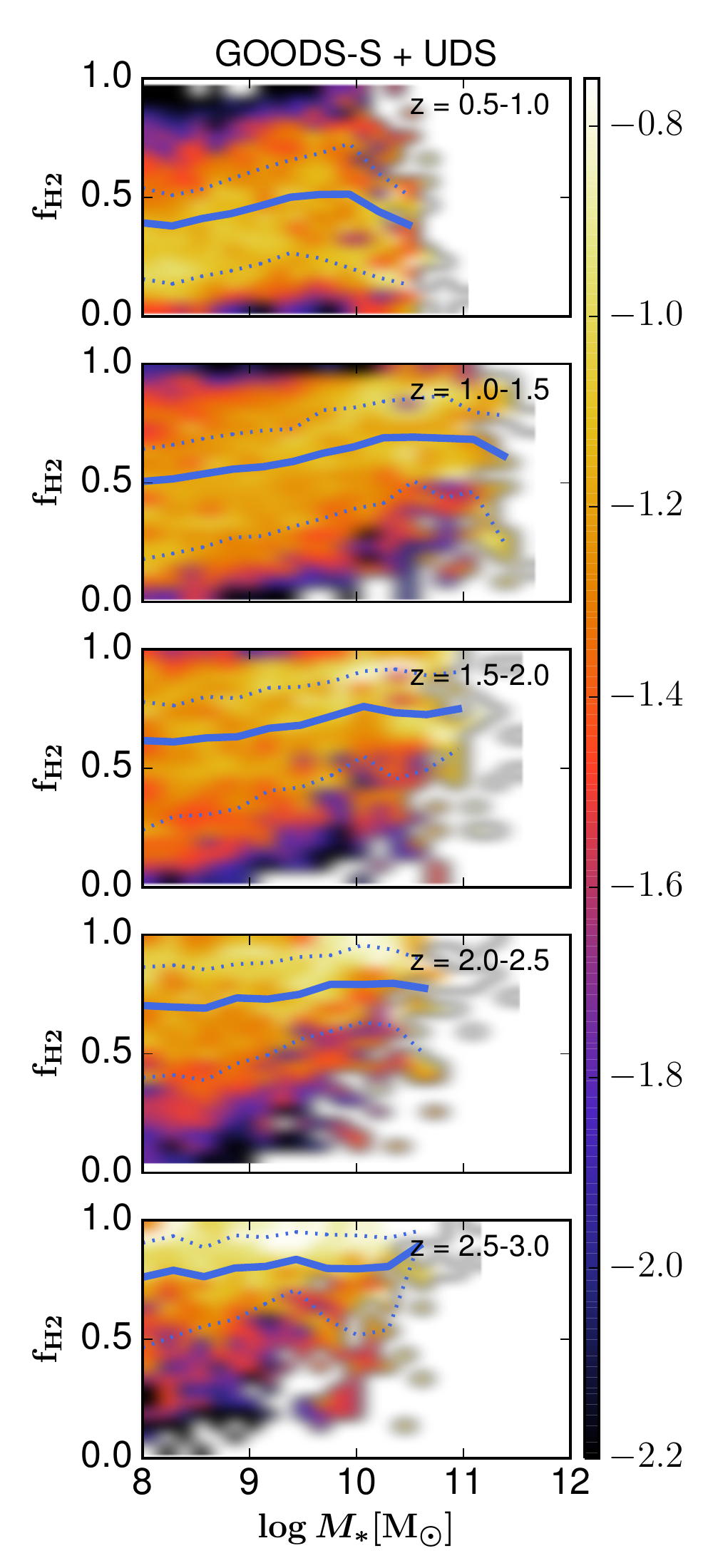}
\caption{Molecular fraction ($\rm{f}_{\rm H2}\equiv M_{\rm H2}/M_{\rm gas}$) of the cold gas as a function of stellar
  mass for different redshift bins. The shaded region shows the
  log of the conditional probability distribution function
  $P(\frac{\rm{M}_{\rm H2}}{\rm{M}_{\rm gas} + \rm{M}_*}|M_*)$, and the blue solid and dotted lines show
  the median fit and one $\sigma$ deviation. 
\label{fig:fh2}}
\end{figure}

\begin{figure}
\includegraphics[width = 1\hsize]{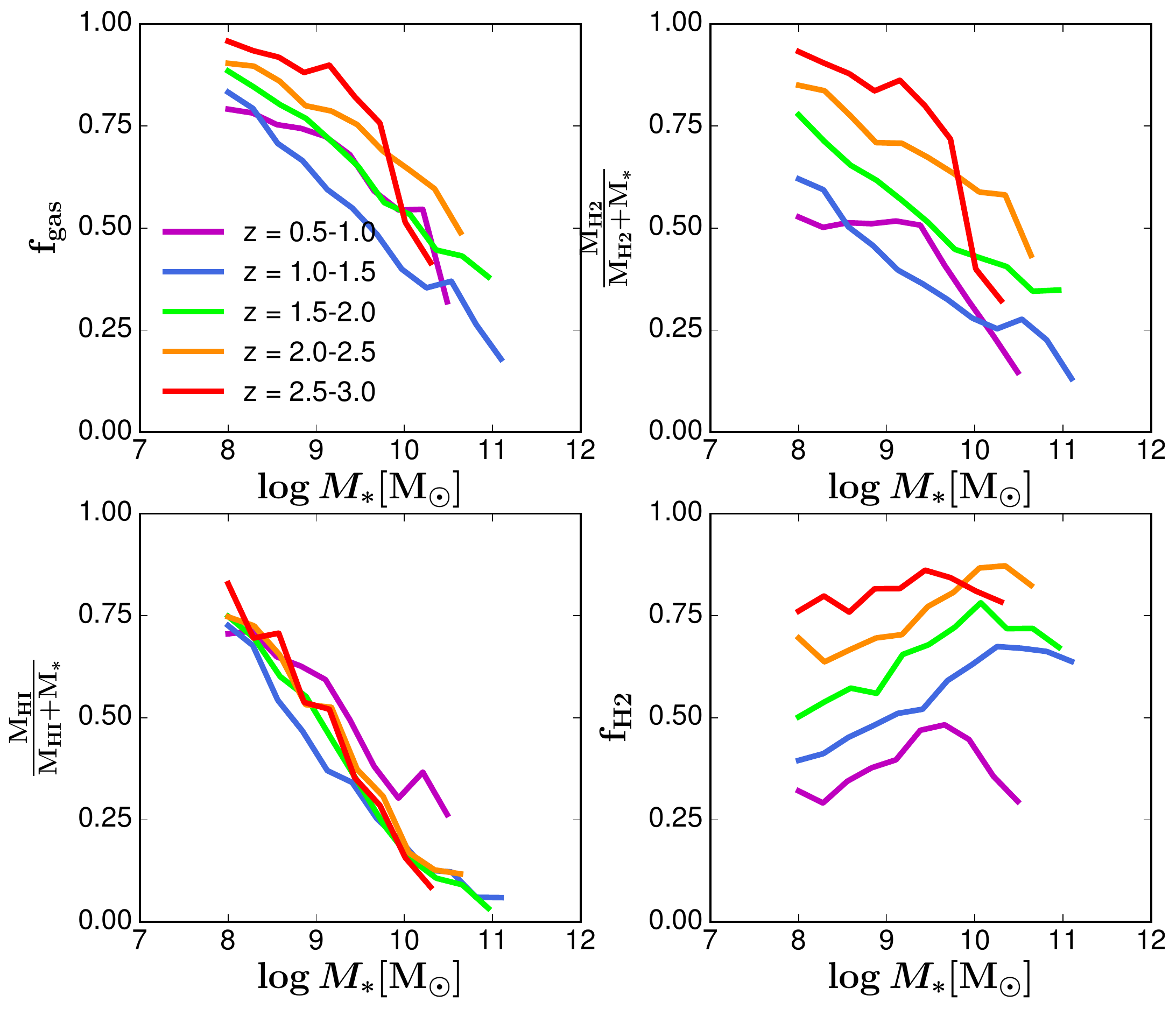}
\caption{Evolution of the mean trends in $f_{\rm gas}$, \frach2star,
\fracHIstar, and $f_{\rm H2}$ as a function of stellar mass. Note that the
mean trend in \fracHIstar with stellar mass remains constant with
time, whereas the other three properties do evolve.
\label{fig:frac_evol}}
\end{figure}

\subsection{Gas fractions}
In Figure \ref{fig:fgas} we show galaxy gas fractions as a function of
stellar mass and redshift. There is a strong anti-correlation between
galaxy stellar mass and gas fraction at $z < 2.5$. The cold gas
fraction remains relatively constant at stellar masses $M*
\leq 10^{9.5}\,\rm{M}_\odot$ and drops
rapidly at higher stellar masses for galaxies in our highest redshift
bin ($2.5 < z < 3.0$). In the lower redshift bins the trend between stellar mass and cold gas
fractions is roughly linear, mainly driven by a decrease in the cold gas fractions
of galaxies with stellar masses $10^{9-10}\,M_\odot$. The
characteristic mass above which the gas fractions rapidly drop ($M_* \approx 10^{9.5 - 10}\,\rm{M}_\odot$)
  suggests that some physical process
  prevents the buildup of large gas reservoirs in galaxies with a
  stellar mass above this characteristic mass, and is similar to the
  quenching mass scale \citep{Kauffmann2003}.
We find the strongest evolution in galaxy cold-gas fraction in intermediate mass
galaxies ($10^{9-10}\,M_\odot$). At a fixed stellar mass, a galaxy’s cold-gas fraction decreases with time.

We present \frach2star $=\frac{M_{\rm H2}}{M_{\rm H2} + M_*}$ as a function of galaxy mass in
Figure \ref{fig:fh2_star}. This quantity is often used as an
observable tracer of the gas content of galaxies. Similar to the total cold-gas fraction, there
is an anti-correlation between \h2 and galaxy stellar mass, as well as
redshift. High stellar mass galaxies have on average the lowest
relative \h2 content, and \frach2star decreases with redshift at a fixed
stellar mass. The trend in \frach2star with stellar mass is similar to
that for total cold-gas fraction. 

We compare the results of our model with direct observations (through
CO) of \frach2star in star-forming galaxies from
\citet{Tacconi2010}, \citet{Saintonge2013}, and
\citet{Tacconi2013}. Although direct observations of \frach2star do not lie on top on
the mean trend, they are in good agreement with the most massive and \h2-rich galaxies in
our sample. As discussed when we described Figure \ref{fig:mstar_mh2},
the difference between the mean trend and
  observations is driven by different selection
  criteria. \citet{Tacconi2013} developed a
  formalism to correct gas fractions for this selection effect and
  found gas fractions of $\sim$30 per cent at $z=1-1.5$ for galaxies
  with stellar masses $\sim 10^{11}\rm{M}_\odot$, in reasonable
  agreement with our inferred gas fractions.

We present the relative amount of atomic hydrogen in galaxies (\fracHIstar $=\frac{M_{\rm HI}}{M_{\rm HI} + M_*}$)
as a function of stellar mass for different redshift bins in Figure
\ref{fig:fracHIstar}. Similar to the previous figures, \fracHIstar
decreases with stellar mass. Below stellar masses of
$\sim\,10^9\,\rm{M}_\odot$ the \hi mass of galaxies exceeds the
stellar mass, whereas at higher stellar mass the \hi mass is
less than the stellar mass. Interestingly, there does not seem to be a strong evolution in
the trend between \fracHIstar and stellar mass with time. A similar
transition mass from \hi to stellar mass dominated is seen for local
galaxies \citep[e.g.,][]{Cortese2011,Huang2012}. 

In Figure \ref{fig:fh2} we present the molecular fraction of the cold
gas ($\rm{f}_{\rm H2}\equiv M_{\rm H2}/M_{\rm gas}$) as a function of stellar mass for different redshift bins. This is a good measure of the fraction of gas available
for SF as a function of time. The
relation between $\rm{f}_{\rm H2}$ and stellar mass is relatively
flat at $z > 2.5$. We find an increasing
trend between $\rm{f}_{\rm H2}$
increasing with stellar mass towards lower redshifts. We find that $\rm{f}_{\rm
  H2}$ increases with stellar mass up to a stellar mass of $\sim
10^{10}\,\rm{M}_\odot$ and decreases at higher stellar masses in our
lowest redshift bins ($0.5 < z < 1.0$).
The average \h2 fraction of
galaxies decreases with time at fixed stellar mass (only at $M_* >
10^{10}\,\rm{M}_\odot$ do galaxies at $2.0 < z < 2.5$
have higher $\rm{f}_{\rm H2}$ than at $z > 2.5$). This result clearly shows that
both the cold gas and \h2 reservoirs decrease with time,
although not necessarily at the same rate. It is important to note
that the relation
between stellar mass and \h2 fraction has a large scatter, especially at low stellar masses and lower redshifts. There is
no well defined drop in \h2 fraction at a characteristic stellar mass
and/or time. 
\begin{figure}
\includegraphics[width = 0.9\hsize]{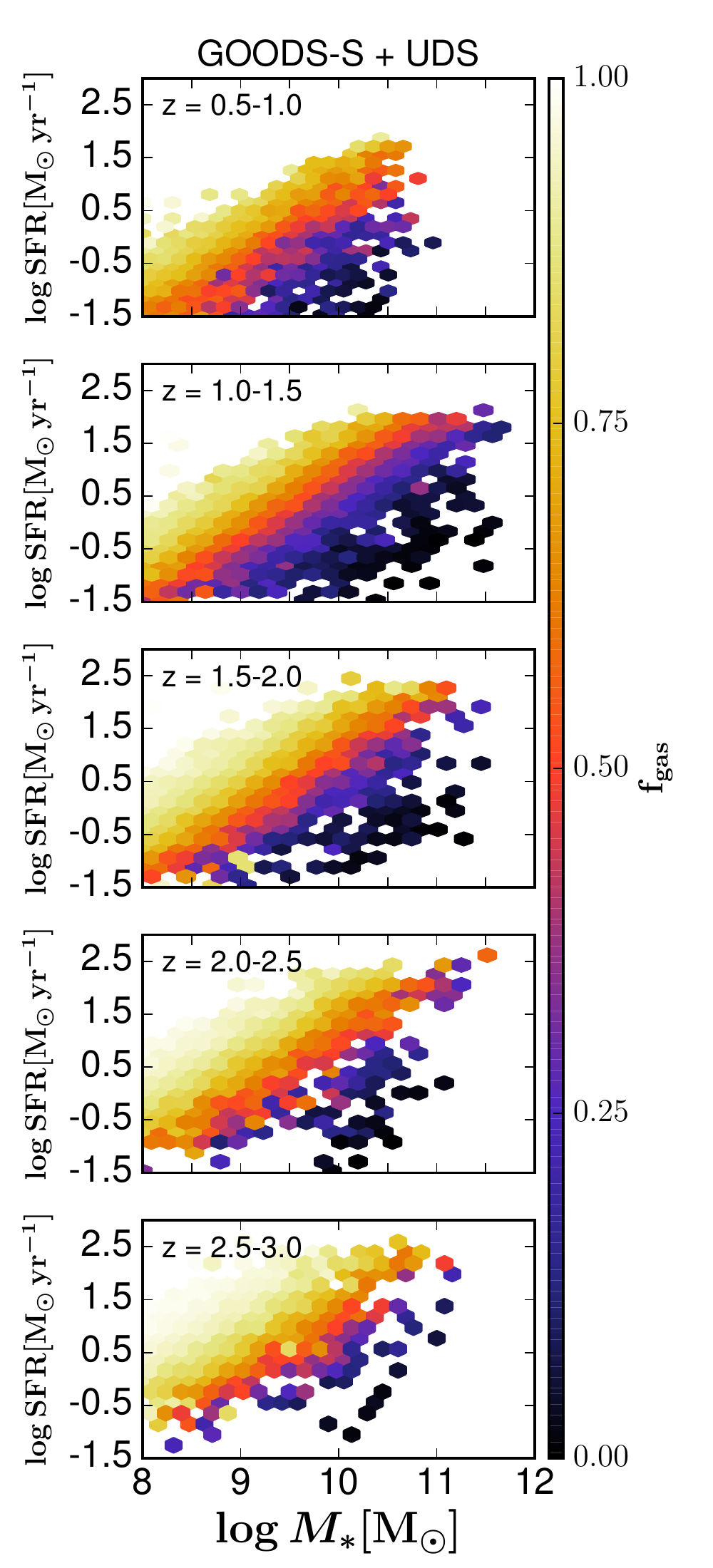}
\caption{Galaxy SFR as a function of stellar mass for different
  redshift bins. The colour map gives the cold gas fraction
  of the galaxies. Each galaxy is counted only
  once to create the hexagons.
\label{fig:mstar.sfr.fgas}}
\end{figure}

\begin{figure}
\includegraphics[width = 0.9\hsize]{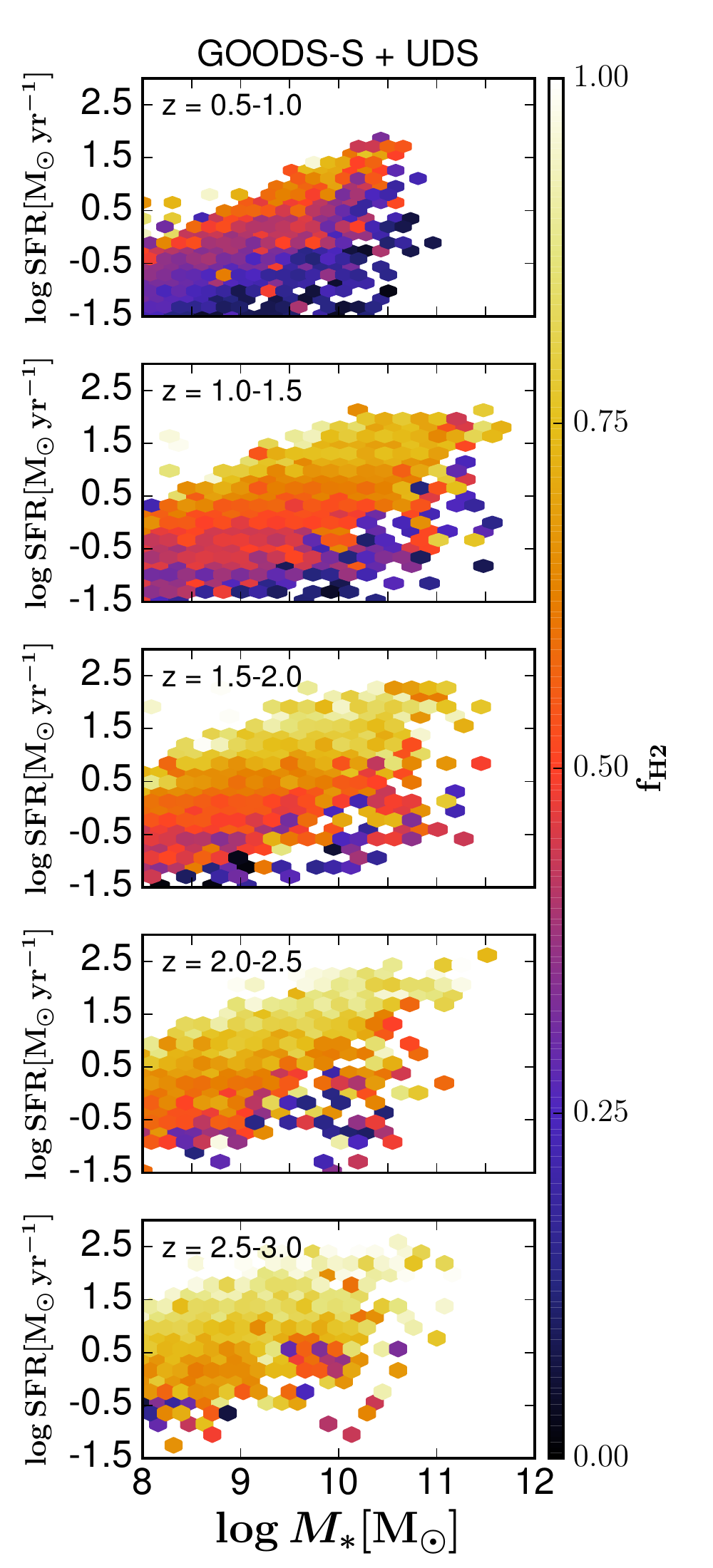}
\caption{Galaxy SFR as a function of stellar mass for different
  redshift bins. The colour map gives the molecular fraction
  of the cold gas. 
\label{fig:mstar.sfr.fh2}}
\end{figure}

We show the mean trends of the previous four Figures again in Figure
\ref{fig:frac_evol}, with the aim of focusing on the
evolution of these trends. We find the remarkable result that $f_{\rm
  gas}$, \frach2star, and $f_{\rm H2}$ all decrease as a function of
time, whereas \fracHIstar remains nearly constant. Although the gas
content of galaxies at fixed stellar mass decreases as a function of
time, the changed partitioning of cold gas in \hi and \h2 results in a constant fraction of \hi in galaxies with
fixed stellar mass. We discuss this further in Section \ref{sec:discussion}.

The CANDELS results
suggest a gradual evolution in the cold gas fraction and the
relative molecular hydrogen content of galaxies at $0.5 < z < 3$, independent of stellar mass.

\subsubsection{Model comparison}
As a comparison we included predictions
from the semi-analytic model presented in
\citet{Popping2013} in Figures \ref{fig:fgas}, \ref{fig:fh2_star}, and \ref{fig:fracHIstar}. We
ran the models in the appropriate redshift bins and applied a
CANDELS selection criteria to the model output ($H_{\rm AB} <
26$).  We only selected disc-dominated galaxies (bulge-to-total ratio
 less than 0.4). 

Although the results presented in this paper are
merely model predictions, they are based on actual observables rather
than introducing uncertain recipes describing the physics
acting on galaxies. A comparison between the results presented in this
work and semi-analytic predictions can shed light on the
successes and failures of theoretical models. We acknowledge that
the results presented are effectively a different representation of the
phase space described by stellar mass, SFR, and galaxy size. One could
therefore also compare the observed values for these quantities to
semi-analytic model predictions. Nevertheless, focusing on the gas
masses of galaxies and potential discrepancies between the results
presented in this work and semi-analytic model predictions may provide a more direct hint of the baryonic
physics that need to be changed and/or included for models to reach
better agreement with observations. \citet{Somerville2014} showed that
semi-analytic models and hydrodynamic models predict too little SF in
galaxies at redshifts $z=1$ and $z=2$ \citep[for a potential solution
see][]{Henriques2014,White2015}. Our method has the potential to
assess whether this is due to too low gas masses in general or too little
\h2.

The semi-analytic model predictions are in good qualitative agreement
with our inferred cold gas fractions at redshifts $z=0.5$ to $z=2.5$
(Figure \ref{fig:fgas}). We predict up to 10 per cent more cold gas in
galaxies with stellar masses larger than $10^{10}\,\rm{M}_\odot$ than
the semi-analytic model. At higher
redshift the semi-analytic model predicts gas fractions higher
than found in this work for galaxies with stellar masses larger than $10^{10}\,\rm{M}_\odot$. 

Semi-analytic model predictions are in
good qualitative agreement with our inferred \h2 masses at redshifts
$1 < z < 1.5$ (Figure \ref{fig:fh2_star}). At lower redshifts the
semi-analytic model predicts much less molecular hydrogen than the
results presented in this paper,
whereas the model predict on average 1.8 times less molecular hydrogen at higher redshifts (\frach2star is
approximately 10 per cent less). In the highest-redshift bin the
model overpredicts \frach2star compared to the
inferred fractions at $M_* > 10^{10}\,\rm{M}_\odot$, whereas it underpredicts the relative amount of \h2 in galaxies at lower
stellar masses. 

The Popping et al. semi-analytic model on average predicts higher values
for \fracHIstar at stellar masses less than $10^{10}\,\rm{M}_\odot$
than the results presented in this work. The model also finds that
at redshifts $z=3-1$ the amount of \hi in galaxies at fixed stellar
mass remains relatively constant. We will further discuss the
implications of this comparison in the discussion.

\subsection{Gas properties on the stellar mass -- SFR diagram}
\label{sec:mstar.sfr.gas}
In Figure \ref{fig:mstar.sfr.fgas} we show the cold gas fraction of galaxies
on the stellar mass--SFR relation. We find that to first order, galaxy
cold gas fractions
are well parameterized by their stellar mass and SFR. It is important to
realize that both the SFR and the stellar mass of a galaxy are key
ingredients in our model that set a galaxy's cold gas mass and its partitioning into atomic and
molecular hydrogen. Galaxy gas fractions stay constant parallel to the relation between SFR and stellar mass, increasing
towards higher SFR and decreasing with stellar mass. The
upper envelope in the stellar mass--SFR diagram is populated by the
most gas-rich objects at a given stellar mass. Although
counterintuitive, galaxies with the highest SFRs do not
necessarily have the highest gas fractions. The highest gas fractions are
found in galaxies with low stellar
masses and high SFRs. Galaxies with high stellar mass and low SFRs have the lowest gas fractions. At a fixed
position in the stellar mass--SFR diagram galaxy cold gas fractions
decrease with time.

We explore the \h2 fraction of galaxies in the stellar mass--SFR
plane in Figure \ref{fig:mstar.sfr.fh2}. Molecular gas fractions
increase with increasing SFR. The most actively star-forming galaxies have
highest molecular gas fractions, and the least-actively star-forming
galaxies have the lowest fractions. Molecular
fractions cannot be parameterized by a simple combination of stellar mass and
SFR as the total cold-gas fractions.

More information can be obtained by comparing Figures
\ref{fig:mstar.sfr.fgas} and \ref{fig:mstar.sfr.fh2}. These figures
clearly show that large \emph{cold}-gas reservoirs do not necessarily lead to
large \emph{molecular}-gas reservoirs and active SF. This is especially evident
for galaxies with low stellar mass ($\sim10^8\,\rm{M}_\odot$) and SFR
$\sim 1\, \rm{M}_\odot \,\rm{yr}^{-1}$ at $z < 1.5$. These galaxies
are dominated by atomic gas. Despite being rich in cold gas, only
approximately 25 per cent of their cold gas is molecular and available
to form stars. These objects were below the observation limit of the
galaxies in \citet{Popping2012}. The deep near-IR photometry
of the CANDELS survey allows us to study these atomic-gas-dominated
low-mass galaxies. Observations of local galaxies show a similar
decrease in \h2 fraction with decreasing stellar mass
\citep{Saintonge2011,Boselli2014,Bothwell2014}, with on average even lower \h2 fractions than the
galaxies in our sample.

The most actively star forming galaxies  ($\sim10^{11}\,\rm{M}_\odot$ and $\rm{SFR} \sim
100\,\rm{M}_\odot \,\rm{yr}^{-1}$) at $z < 2.5$ have relatively small
cold gas reservoirs ($\rm{f}_{\rm gas} < 0.5$), but most of this cold
gas is molecular. Galaxies with high stellar mass and low SFRs are nearly
empty of cold gas, but the molecular fraction of this gas is
approximately 0.5 and can account for some residual star-formation.

\begin{figure}
\includegraphics[width = 0.9\hsize]{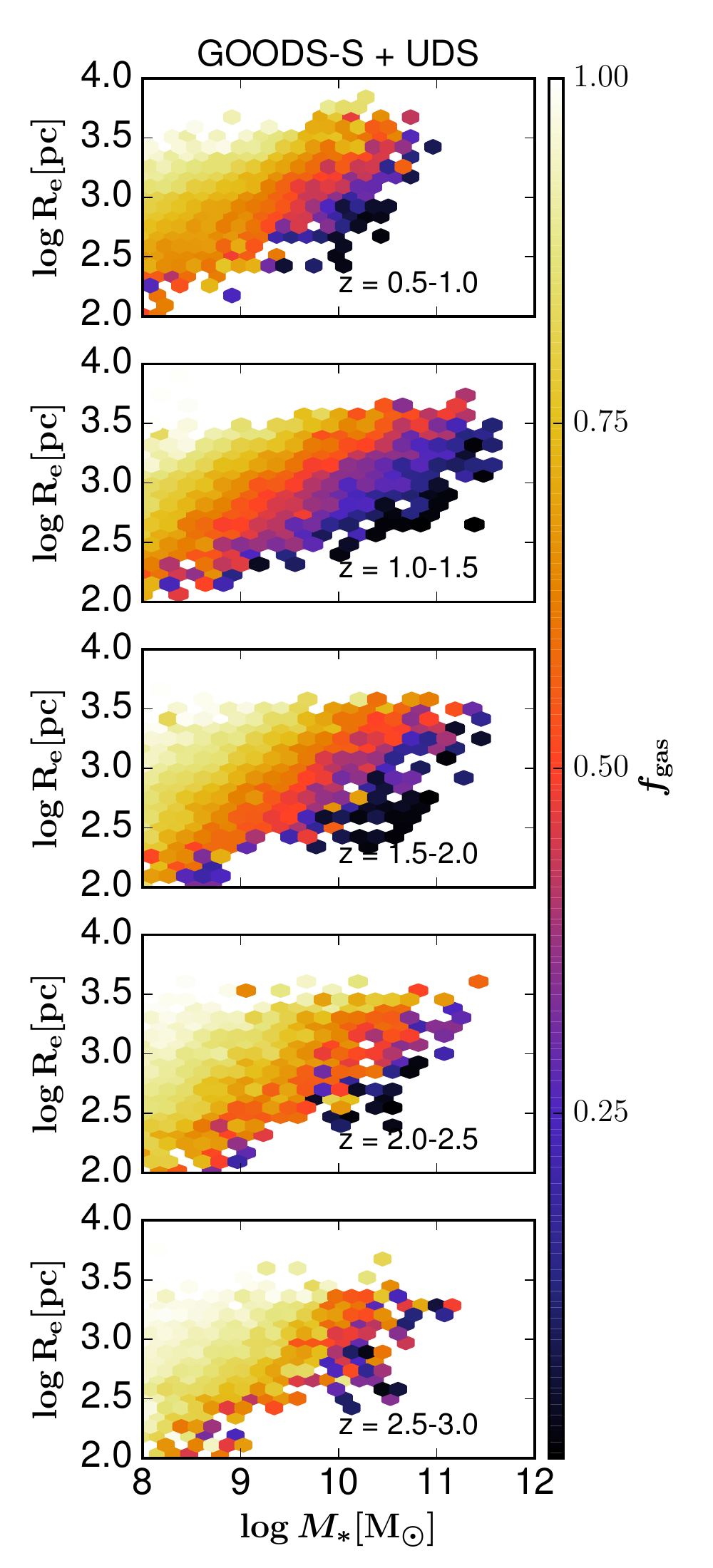}
\caption{Galaxy scale radius as a function of stellar mass for different
  redshift bins. The colour map gives the cold gas fraction. 
\label{fig:size_gas}}
\end{figure}
\begin{figure}
\includegraphics[width = 0.9\hsize]{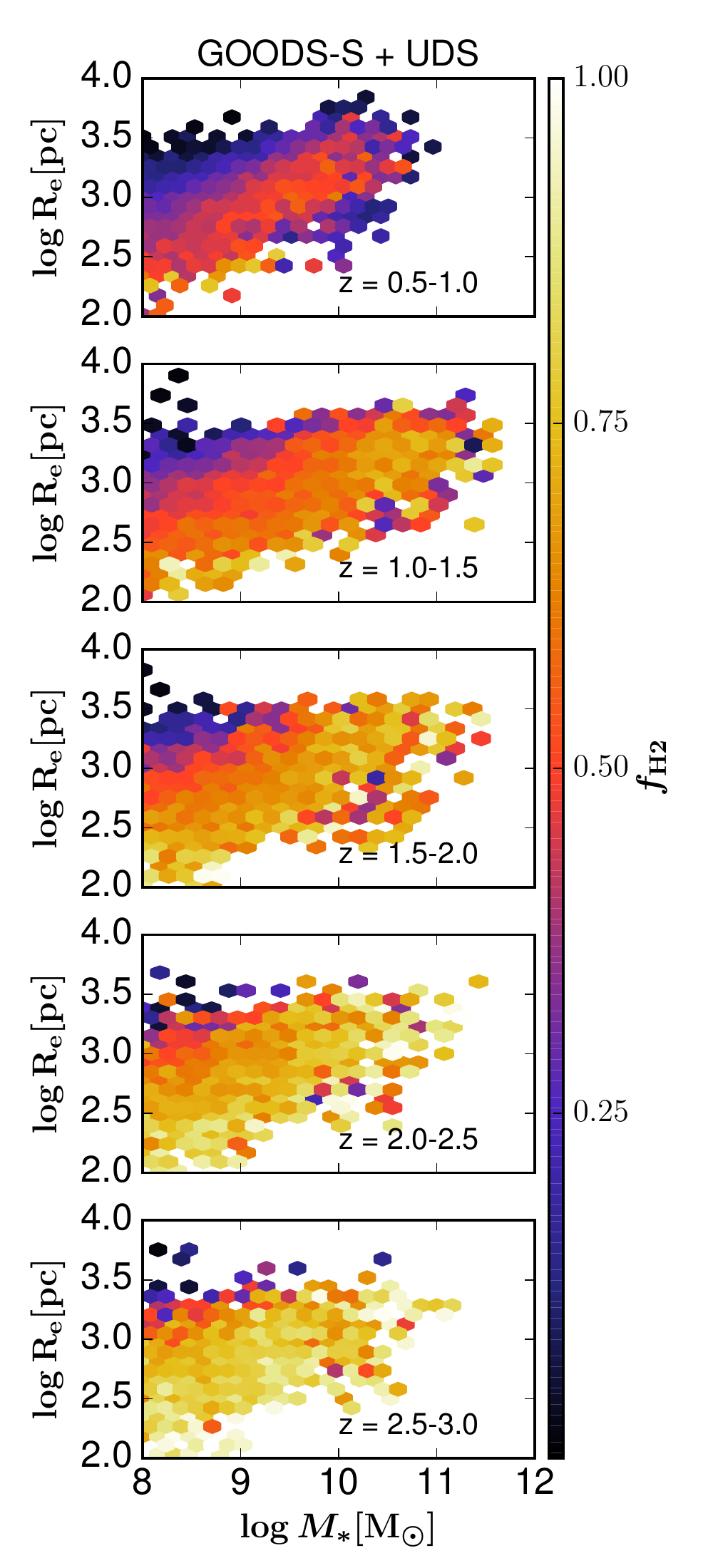}
\caption{Galaxy scale radius as a function of stellar mass for different
  redshift bins. The colour map gives the molecular fraction
  of the cold gas. At fixed stellar mass more
  compact galaxies have higher molecular fractions.
\label{fig:size_H2}}
\end{figure}

\begin{figure}
\includegraphics[width = 0.9\hsize]{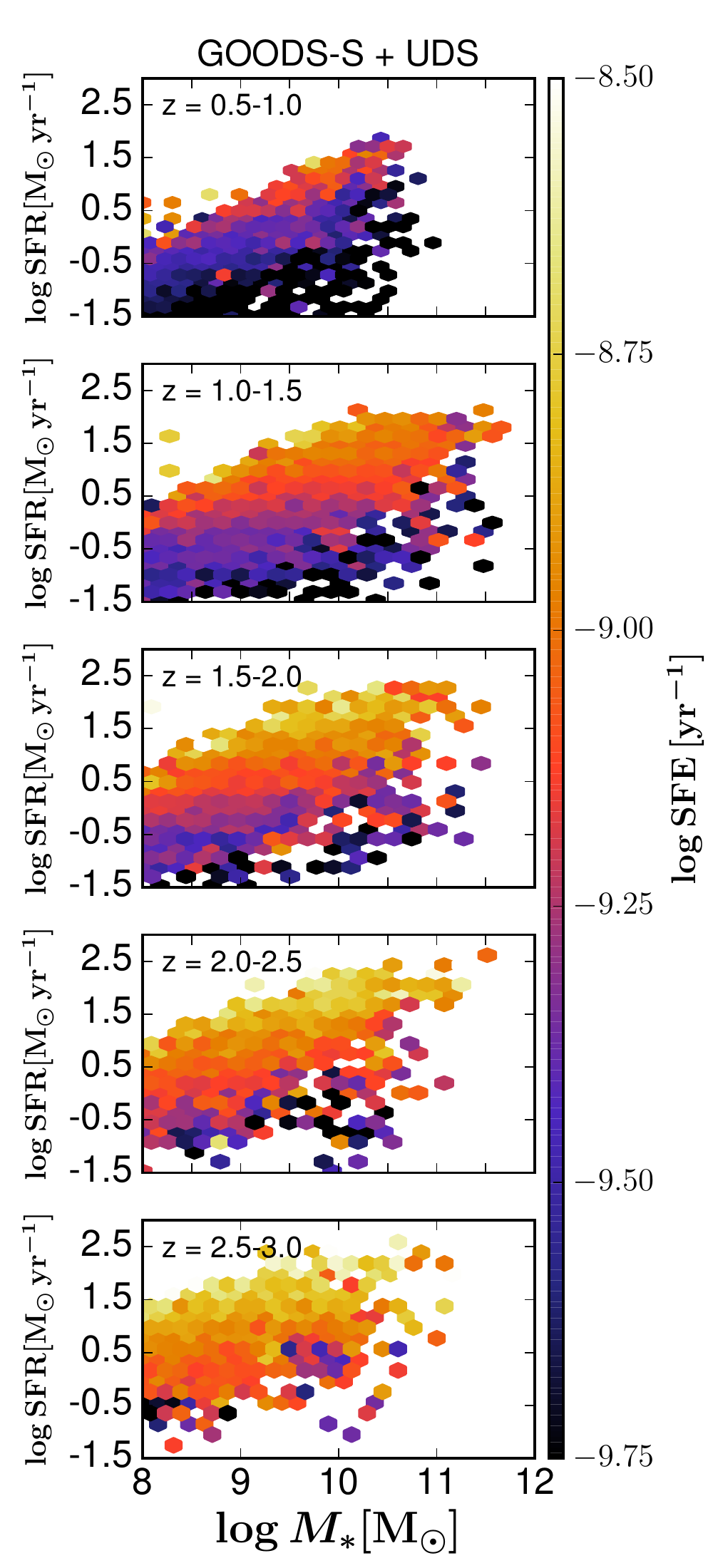}
\caption{Galaxy SFR as a function of stellar mass for different
  redshift bins. The colour map gives the star-formation
  efficiency ($\rm{SFE} \equiv {\rm SFR}/M_{\rm
  gas}$) of the galaxies. 
\label{fig:sfr_SFE}}
\end{figure}
\begin{figure}
\includegraphics[width = 0.9\hsize]{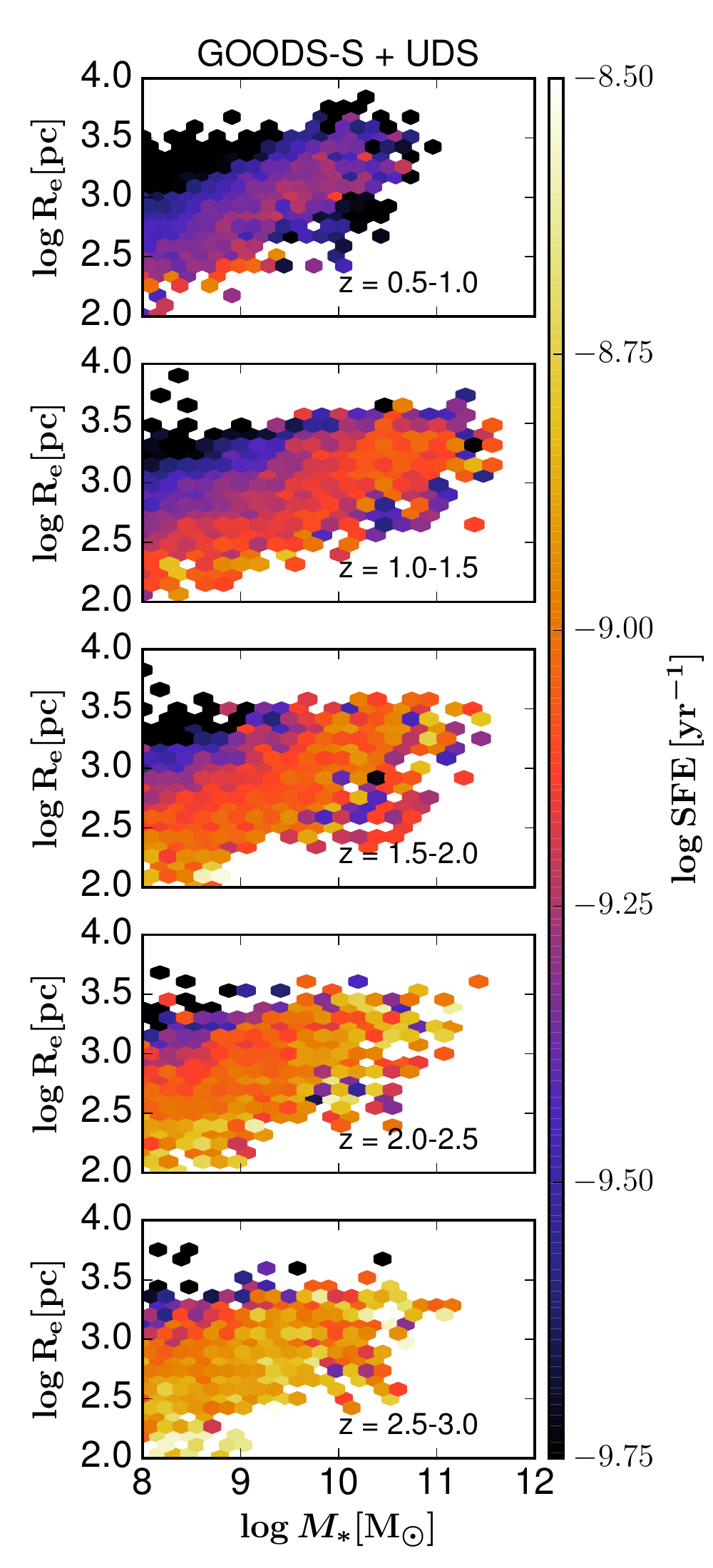}
\caption{Galaxy scale radius as a function of stellar mass for different
  redshift bins. The colour map gives the star-formation
  efficiency ($\rm{SFE} \equiv {\rm SFR}/M_{\rm
  gas}$) of the galaxies. 
\label{fig:size_SFE}}
\end{figure}
\subsection{Gas properties on the stellar mass -- size relation}
We present the cold-gas and molecular fraction of galaxies on the stellar mass--size
relation in Figures \ref{fig:size_gas} and \ref{fig:size_H2}. Within
our model the scale radius of a galaxy sets the surface densities that
control the molecular
fraction and the total mass of cold gas in that galaxy. At fixed stellar mass the cold gas
fractions of star-forming galaxies increase with increasing size. The molecular fraction of the cold gas, on the other hand,
decreases with increasing size. For a fixed scale radius the cold-gas fraction and molecular
fraction of a galaxy decrease with time.
 
The largest variations in molecular
fractions are found in low-stellar-mass galaxies ($M_* <
10^{10}\,M_\odot$). At higher stellar masses we find differences in
molecular fraction of approximately 25 per cent. Although less
dramatic than for lower-mass galaxies, such variations are still
significant.

\subsection{Star-formation efficiencies}
The star-formation efficiency (the ratio between SFR and total cold
gas mass (\hi + \h2), i.e., $\rm{SFE} \equiv {\rm SFR}/M_{\rm
  gas}$) is a good measure of how efficiently galaxies turn cold gas
into stars.\footnote{We note that the star-formation efficiency
  of galaxies is often defined as the ratio between SFR and molecular
  hydrogen mass \h2 (i.e., $\rm{SFE}\equiv {\rm SFR}/M_{\rm H2}$). Our
  adopted definition of the ratio between SFR and total cold gas mass
  is slightly different and important for simulations that do not
  discriminate between \hi and \h2.} We show the SFE of galaxies on the stellar mass--SFR and stellar mass--size diagrams in Figures \ref{fig:sfr_SFE} and
\ref{fig:size_SFE}. 

We find that galaxy SFEs increase with
increasing SFR and decrease with time. The SFE of galaxies that
populate the upper envelope in the stellar mass--SFR relation changes as a
function of stellar mass. At high stellar masses, galaxies are more
than twice as efficient in forming stars out of their total cold gas
reservoirs (\hi + \h2)
than at low stellar masses. This shows that the position of a galaxy
on the stellar mass--SFR plane is being driven by the
amount of cold gas available \emph{and} by the ability of the gas to
form molecules and stars \citep{Saintonge2012,Genzel2014,Sargent2014}. The group of galaxies with low SFRs at high stellar masses is
more than twice as inefficient in forming stars than their
counterparts with the same stellar mass that are actively forming
stars. Overall, the SFE of galaxies
decreases with time. 

The SFE of galaxies decreases with increasing scale radius. This is
especially clear for galaxies with low stellar mass, but is also apparent for
the most massive galaxies in our sample. This result is in good
agreement with our findings in Figure \ref{fig:size_H2}. Compact galaxies
have higher molecular fractions and are therefore more efficient at forming stars.

\subsection{Global evolution of cold gas content of galaxies}
The evolution of \frach2star with redshift has been studied in several
different CO surveys and through dust continuum estimates \citep{Geach2011,Tacconi2010,Daddi2010,Magdis2012,
  Tacconi2013,Saintonge2013,Bauermeister2013, Genzel2014,Scoville2014}. These surveys suggest
\frach2star increases with increasing redshift up to $z\sim2$ and
flattens or even declines at higher redshifts \citep{Saintonge2013}.

\begin{figure}
\includegraphics[width = 1.0\hsize]{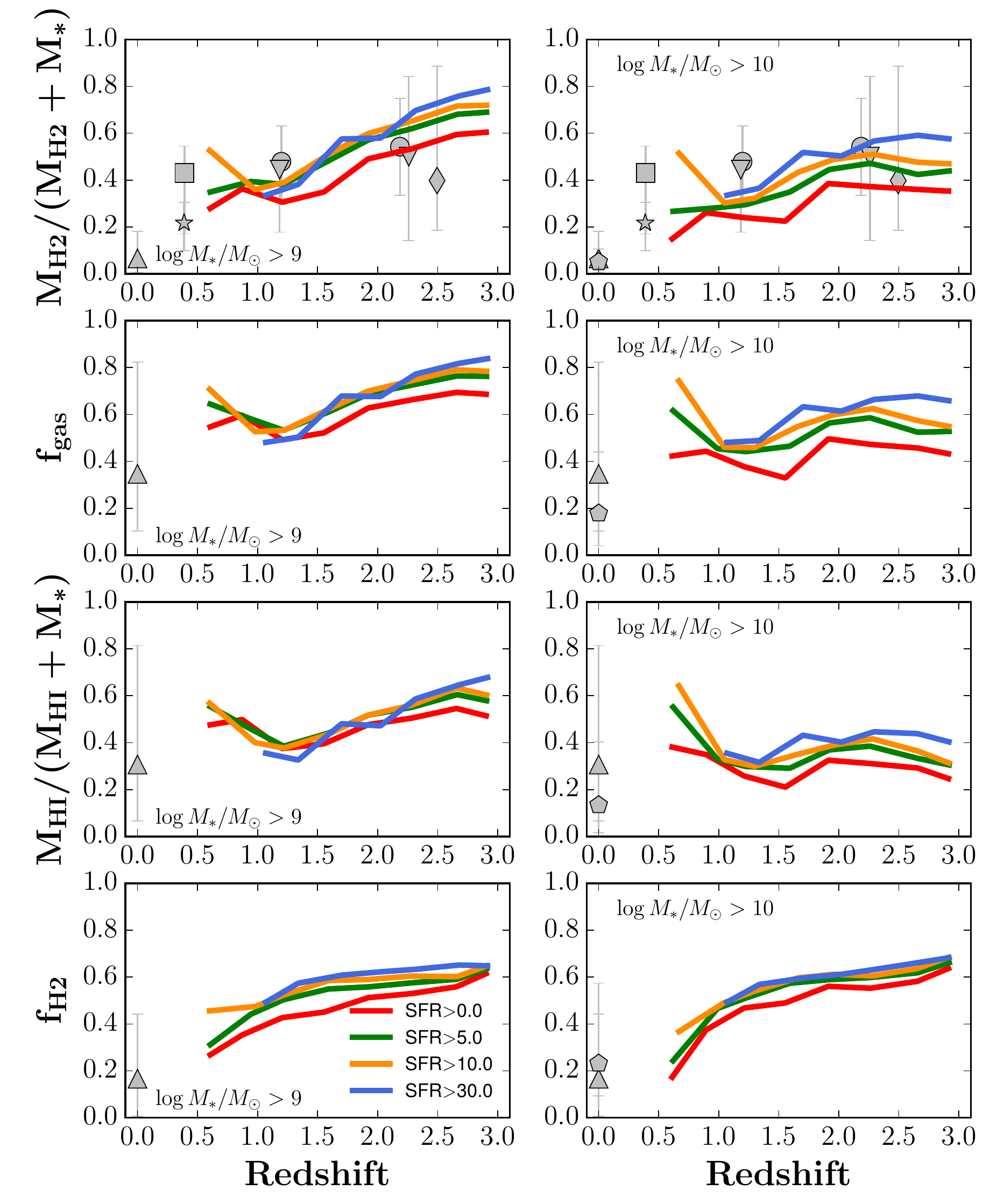}
\caption{Redshift evolution of the  relative \h2 content
  $\frac{M_{\rm{H2}}}{M_{\rm{H2}} + M_*}$ (top row), gas fraction
  (second row), relative \hi content (third row), and molecular hydrogen fraction (bottom row) for galaxies with stellar
  masses in the range $M_* > 10^{9}\,M_\odot$ (left panels), $M_* >
  10^{10}\,M_\odot$ (right panels), and SFR larger than 0
  (red), 5 (green), 10 (orange), and 30 (blue) $\rm{M}_\odot\,\rm{yr}^{-1}$. Different
markers represent the following datasets: {\em upwards pointing triangles}: galaxies from the
THINGS survey \citep{Leroy2008} at $z=0.0$, {\em pentagons}: galaxies from the
COLD GASS survey \citep{Saintonge2011} at $z=0.0$, {\em squares}:
galaxies from \citet{Bauermeister2013} at $z\sim 0.4$, {\em stars}:
galaxies from \citet{Geach2011} at $z\sim 0.4$, {\em circles}: galaxies
from the PHIBBS survey \citep{Tacconi2013} at $z\sim 1.2$ and $z\sim
2.2$, {\em downwards pointing triangles}: galaxies at $z\sim 1.2$ and $z\sim
2.2$ from \citet{Tacconi2010}, {\em diamonds}: galaxies at $z\sim 2.2$ from
\citep{Saintonge2013}. Error bars show the 80 percent confidence
interval for each observational sample at the given redshift bin.
\label{fig:fh2_z}}
\end{figure}

We present the evolution in \frach2star for galaxies with
different selection criteria in stellar mass and SFR in
Fig. \ref{fig:fh2_z} (top row). Our method allows us to
  infer the gas content for a large number of galaxies covering a wide
  range in physical properties (stellar masses, SFRs). This makes it
  ideal to investigate the effects that different selection criteria
  have on cold gas evolution trends inferred from direct observations
  of the CO emission line and the dust continuum of galaxies.
We find a decrease in the relative \h2 content
of galaxies from $z\sim 3$ to $z\sim 1$. There is a minor increase in
the relative \h2 content (and other fractions) from $z=1.0$ to
$z=0.5$ when selecting galaxies with stellar masses larger than $10^9\,\rm{M}_\odot$. Above $z = 2.0$ the
evolution of \frach2star becomes less steep. This observed
evolution is similar to the results presented in \citet{Saintonge2013}. The normalization and the shape
of the evolution in \frach2star depend on the
selection criteria applied. We find the best match with the current direct observations when
selecting only galaxies with ${\rm SFR} > 30\,M_\odot\,\rm{yr}^{-1}$ for galaxies with
stellar masses $\log{({\rm M}_*\,/{\rm
    M}_\odot)}> 10.0$, similar to the selection limits of the observed
samples. When loosening the selection criteria for
SFR the normalization of \frach2star decreases. These galaxies with
low SFRs are currently  typically not accounted for in direct surveys of the molecular content
of galaxies through either CO or FIR sub-mm continuum studies. When including galaxies with lower stellar masses we find that not only the normalization in the evolution
of \frach2star increases, the slope of the decline in \frach2star at $z\leq2$
also becomes stronger. We find no changes in the evolution of
\frach2star when selecting galaxies with stellar masses more massive
than $10^{9}$ solar masses and changing the SFR criteria.

The relative amount of atomic hydrogen in galaxies slightly decreases
with time in galaxies with stellar masses larger than
$10^9\,\rm{M}_\odot$ and remains constant in galaxies with stellar masses more
massive than $10^{10}\,\rm{M}_\odot$. 

The mean molecular fraction $f_{H2}$ of the cold gas in galaxies
gradually decreases over the entire redshift range probed. This is in good agreement with the results presented
in Figure \ref{fig:fh2} where we showed that the mean molecular fractions as a
function of stellar mass decreases by approximately 25 per cent for
the lowest mass galaxies from redshift 3 to 0.

\section{Model assumptions and applicability to the CANDELS sample}
\label{sec:assumptions}
The presented approach is based on several assumptions that may not
always hold. In this section we discuss the main assumptions and
how they affect our inferred gas masses. 

We converted all the sizes to a common restframe wavelength of 5000
{\AA}. Originally they were based on $H$ band
photometry. Across the $z=0.5$--$3$ redshift range the $H$-band
photometry spans a restframe size from the $B$ band to the $I$ band. 
Inferred gas masses based on
the 5000 {\AA} restframe morphology are slightly higher than based on
the original $H$-band morphology. When adopting the 5000 {\AA}
restframe morphology, less than 5\% of the galaxy sample is more than
10\% more gas
rich than when adopting the $H$-band morphology. Only a handful of
objects is more than 20\% more gas rich when adopting a 5000 {\AA}
morphology, rather than the $H$-band morphology.

A key assumption in our method is an exponential distribution of
stars and gas in the galaxy discs. SF takes place in molecular clouds
(local clumps in the disc not following an exponential distribution)
which could lead to a local underestimation of the cold gas surface
density. In this work we are averaging over $\sim$kpc scales and at our
redshifts of interest and at these scales the exponential disc model is a good approximation to the stellar distribution of star-forming galaxies on galactic scales \citep{Wuyts2011}. Furthermore, analyses of local galaxies have
revealed an exponential distribution of the cold gas in star-forming
galaxies \citep{Leroy2008,Bigiel2012,Kravtsov2013, Wang2014}. 

Our model assumes a constant ratio between the gas and stellar
disc scale lengths ($\chi_{\rm gas}=1.7$), based on fits to the
galaxy disc profiles in \citet{Leroy2008}. This number is in good
agreement with typical ratios of 1.5--2.0 between \hi and optical disk
sizes \citep{Verheijen2001}. \citet{Kravtsov2013} finds
a $\chi_{\rm gas} = 2.6$, when normalizing gas
profiles to a radius $r_n$ obtained using abundance matching
arguments. Furthermore, the value for $\chi_{\rm gas} $ may vary as a function of redshift or galaxy properties. Varying $\chi_{\rm gas}$
results in negligible differences in inferred cold gas and \h2 mass as
long as $\chi_{\rm gas} > 1$. Decreasing $\chi_{\rm gas}$ to values
less than one leads to more significant differences in the inferred
gas masses. When adopting the extreme case of $\chi_{\rm gas} = 0.5$ we find that the
inferred gas and molecular masses are lowered by 0.25 and 0.1 dex,
respectively. Observations of the sizes of the CO discs at
$z\sim1.2$ and $z\sim2.2$ (supposedly tracing the molecular hydrogen)
do not support a scale length ratio of $\chi_{\rm gas} < 1.0$
\citep[e.g.][]{Tacconi2010,Tacconi2013}. \citet{Berry2013} found that
in semi-analytic models \hi discs have to be more extended than the
stars in order to reproduce Damped Lyman-alpha properties \citep{Schaye2001}.

A key ingredient in our model is the partitioning of cold gas into a
molecular and atomic component using a pressure-based recipe presented
in \citet{Blitz2006}. This recipe was calibrated based on observations
in local galaxies \citep{Blitz2006,Leroy2008} and has not yet been constrained
at our redshifts of interest. An increased UV background or lower gas metallicities could suppress the formation of molecules on dust grains
and the ability of the molecular hydrogen to
self-shield. Under these conditions our model would require a larger
inferred cold gas reservoir to support the observed SFRs. We
anticipate that future high-resolution observations with ALMA and SKA
will be able to assess the validity of these locally calibrated
relations at high redshifts. An alternative approach would be to use a
molecular hydrogen prescription that is also a function of gas phase
metallicity \citep[e.g.,][]{Krumholz2009,Gnedin2011}. Similarly to the
pressure based algorithm, these
prescriptions have also only been tested against local observations
\citep[e.g.][]{Fumagalli2010}. A significant downside
of applying this approach is that gas-phase metallicities are only
available for a limited number of galaxies at our redshifts of interest. A different approach could be
to use the fundamental metallicity relation \citep{Mannucci2009} to
estimate the gas-phase metallicity of galaxies as a function of their
SFR and stellar mass. We opted not to use this approach, as this will introduce an
additional empirical step and source of uncertainty. Moreover, the
calibration of gas phase metallicity measurements is highly uncertain.

We assume an adapted version of the SF relation presented in
\citet{Bigiel2008} to allow for an increased star formation efficiency
in high-surface-density environments. When not accounting for the
increased star-formation efficiency (i.e. $N_{\rm SF} = 0$), the
presented method predicts larger values for the gas content of galaxies with
$\rm{SFR}/(2\pi r_*^2) \approx 1\,\mathrm{M_\odot yr^{-1} kpc^{-2}}$ by
roughly 10 percent and $\rm{SFR}/(2\pi r_*^2) \approx
10\,\mathrm{M_\odot yr^{-1} kpc^{-2}}$ by roughly 75 percent (less
than 1 percent of our sample has a SFR surface density larger than
$\rm{SFR}/(2\pi r_*^2) \approx 10\,\mathrm{M_\odot yr^{-1}
  kpc^{-2}}$).

To estimate the mid-plane pressure on the disc we adopt the
prescription presented in \citet{Elmegreen1989}. We estimate the ratio
between the gas and stellar vertical velocity dispersion following
\citet{Fu2010}, based on empirical scalings for nearby disc
galaxies. Observations suggest that star-forming galaxies at $z\sim2$
are ``puffy'' with high velocity dispersions
\citep{Foerster2009}. Depending on the actual distributions of the
giant molecular clouds and star-forming clumps, this may
change the shielding properties of the molecular hydrogen.

An additional source of uncertainty lies in the conversion from CO
emission to \h2 mass \citep[see][for a review]{Bolatto2013} or the
conversion from sub-mm dust continuum to gas mass. Our method is
calibrated on \h2 masses that rely on the aforementioned
conversions. Our method therefore includes these
uncertainties. 

In \citep[e.g.,][, Figure 3]{Popping2012} we showed that propagating the
uncertainty in the input parameters (stellar mass, SFR, and size), the typical
systematic uncertainty in inferred gas masses is approximately
0.3 dex. Many of the uncertainties cannot be well quantified. Combining all the concerns discussed above as well as the concerns
discussed in \citet[Section 3.3]{Popping2012}, and taking the errors in
the input parameters into account, we estimate an uncertainty in the inferred gas masses for our method is approximately 0.7 dex.

\section{Discussion}
\label{sec:discussion}
\subsection{Galaxy gas fractions}
The gas fraction of a galaxy is set by the
competition between the inflow, outflow, and consumption of cold gas in
the galaxy \citep{Bouche2010,Dave2011,Dekel2014}. The cosmic SFR
peaks at $z\sim 2$ \citep{Hopkins2006,Madau2014}, and analytic calculations and cosmological hydrodynamical
simulations predict large gas accretion rates onto galaxies at this
epoch \citep{Birnboim2003,Keres2005,Keres2009,Dekel2006,Dekel2009}. Our indirect gas measures
provide an ideal sample to constrain which of the aforementioned
processes dominate this competition at different cosmic times. We find
that gas fractions decrease most rapidly at $z < 1.5$, although a
gradual decrease is present at all redshifts studied in this sample.
These results suggest that the
accretion of gas onto galaxies cannot keep up with the consumption
and/or outflow of gas, especially at redshifts $z<2$ \citep[see
also][]{Popping2012}. These results are in good qualitative agreement
with predictions from semi-analytic and semi-empirical models  \citep[e.g.,][]{Obreschkow2009,Lagos2011cosmic_evol,Fu2012,Popping2013,Popping2015}. Besides the relative gas content, the molecular fraction of the cold
gas also decreases with time, most rapidly at $z<2$ \citep{Saintonge2013}. The
characteristic density of the cold gas decreases
\citep{Popping2013RT}, allowing less gas to self-shield and become
molecular. This indicates
that the physical process that suppresses star-formation with time is
at least two-fold: galaxies run out of gas as well as molecules, but
not necessarily at the same rate. For the most massive galaxies in our
sample this two-fold process is already taking place at $z\sim3$, well
before the peak in cosmic SFR and before the accretion of cold gas
onto galaxies slows due to the increasing dominance of dark energy
driven expansion. This is an important constraint for
models of galaxy formation that include \h2-based star-formation recipes.

The rapid decrease in cold gas fractions below $z<1.5$ is supported by
direct observations of the CO emission in galaxies \citep[Figure \ref{fig:fh2_z}][]{Tacconi2013,Saintonge2013}. These observations indicate that the relative \h2 content of galaxies remains flat above
redshifts of $z\sim 2$ and gradually decreases at lower
redshifts. The slope and normalization of this trend is in part set by the selection criteria. Depending
on the selection criteria in stellar mass and SFR chosen, the evolution of \frach2star at redshifts
$z>2$ can change. 

\citet{Popping2012} found for a sample of galaxies taken from COSMOS
at redshifts $0.5 < z < 2$  that the cold gas fraction of
galaxies with stellar masses of
$\sim 10^{10.5}\,\rm{M}_\odot$ evolves from $\rm{f}_{\rm gas} \sim 1$
at $1.75 < z < 2$ to $\rm{f}_{\rm gas} \sim 0.2$
at $0.5 < z < 0.75$. The authors found that less-massive galaxies showed hardly any evolution in their gas fractions. This difference suggested that massive galaxies become gas-poor earlier
and quicker than less massive galaxies; this behaviour was characterized as
downsizing in gas content. Furthermore, the authors found a
characteristic stellar mass below which the gas fraction of galaxies
rapidly drops. This characteristic mass decreased towards lower redshifts. We find a similar characteristic
stellar mass at redshift $2.5 < z < 3$, but the drop in the relation
between cold gas mass and stellar mass disappears towards lower
redshifts. Also, there is no strong difference in the rate at which
the highest mass galaxies run out of cold gas compared to the least
massive galaxies in our sample (see Figure \ref{fig:frac_evol}). We remind the reader that the CANDELS
survey allows us to study much fainter objects and is not biased towards the most
star-forming objects. We ascribe the conclusion
reached in \citet{Popping2012} to the selection of only the most
actively star-forming, and therefore most gas rich, galaxies at stellar masses less than
$10^{9.5}\,\rm{M}_\odot$. Our results suggests a gradual evolution in the cold gas fraction and the
relative molecular hydrogen content of galaxies at $0.5 < z < 3$,
independent of stellar mass.

\subsection{Extended parameter space and selection effects}
With the technique used in this work we can easily infer the cold gas
masses for large samples of galaxies, covering a wide range in stellar
mass, SFR, and redshift. As discussed above, this allows us to
evaluate the impact of selection effects on smaller surveys of directly
measured gas masses and to extend observed relations to stellar mass, SFR, and
redshift regimes where gas measure are hard to obtain directly.

CO observations of actively star-forming galaxies have revealed a
linear relation between \h2 mass and stellar mass \citep[e.g.][]{Saintonge2013,Tacconi2013}. Our results
suggest that this relation was already in place at $z=3$, when the
Universe was less than 3 Gyr old and extends down to stellar
  masses of at least $10^{8}\,\rm{M}_\odot$. A similar relation (although less
well defined) is found for the total cold gas mass and \hi mass. The
trend between stellar mass and \hi mass has larger scatter than the trend
between stellar mass and cold gas mass and \h2 mass. Even though the uncertainties in the gas masses
are significant, these extended scaling relations suggest that the
balance between the consumption and in- and outflow of gas in galaxies
is already at place for galaxies with low stellar masses.

In Figure \ref{fig:fh2_z} we explored the direct effect of changing
selection criteria on the evolution of gas fractions in
galaxies. We found no qualitative difference between different
selection criteria. The fraction of gas, \hi, and \h2 in galaxies
decreases gradually from $z=3$ to $z=0.5$.  The absolute values of the
gas fractions change by a factor of a few when changing the selection
criteria. We find that for a sample of galaxies with stellar masses
more massive than $10^9\,\rm{M}_\odot$ the inferred cold gas, \hi, and
\h2 fractions of galaxies remain nearly identical when galaxies with a
reasonable amount of SF are selected (SFRs of a few solar masses per
year). Future surveys with a stellar mass limit of
$10^9\,\rm{M}_\odot$ should be able to robustly constrain the
evolution of the gas content of galaxies contributing most to the
stellar mass budget and stellar growth in our Universe.

We found good agreement between direct observations of the \h2 mass of
galaxies and our inferred \h2 masses for the galaxies with the highest
SFR in our sample (with $\rm{SFR} > 20 -
30\,\rm{M}_\odot\,\rm{yr}^{-1}$). Our results indicate that at the
stellar masses probed by direct surveys, there is also a substantial
population of galaxies with \h2 masses less massive than
$10^{10}\,\rm{M}_\odot$. This group of galaxies should be accounted for
when trying to make more generalized statements, especially when
focusing on galaxies with large stellar masses.

\subsection{A constant relation between \hi mass and stellar mass}
We found hardly any evolution in the relation between \hi mass and
stellar mass, whereas the relation between cold gas mass and \h2
mass with stellar mass does evolve. \citet{Popping2015} found a similar result
when coupling an abundance matching model with the semi-empirical
approach to infer gas masses adopted in this work. Semi-analytic, semi-empirical, and
hydrodynamic models have also found a constant relation between \hi mass and stellar mass remains constant with time (for galaxies with stellar
masses larger than $10^{9}\,\rm{M}_\odot$), just as the \hi mass
function out to $z=2$ \citep{Dutton2010,Lagos2011cosmic_evol,Fu2012,
  Dave2013,Popping2013,Popping2015}.  

Although already predicted by different
models, this is the first time a model used actual observations  as
input to find a constant \hi-to-stellar mass ratio. The relatively weak
evolution in \hi mass is driven by an apparent self-regulation that
naturally arises in galaxies. As gas is consumed over time and
galaxies grow as a function of stellar mass, the cold gas surface
density decreases which naturally causes the \hi fraction of the cold
gas to increase. Although this process may prevent a rapid drop in the
\hi reservoirs of galaxies, by no means does this process imply that
the \hi mass at a fixed stellar mass should remain constant. There is a
fine balance between the amount of cold gas being consumed, heated and/or
expelled, and the partitioning of the remaining gas into an atomic and molecular component
that sets the constant \hi mass. 

We are aware that this result should
be placed within the context of the model framework adopted in this
(and other) work. If the adopted recipes for SF and the partitioning of cold gas are not
valid for galaxies at $z>0$ this could also dramatically change our
conclusions. We therefore look forward to surveys with the SKA
pathfinders probing the \hi mass of galaxies out to at least
$z=1$. If indeed the \hi-to-stellar mass fraction of galaxies remains
constant, it should be feasible to observe the \hi mass of $M_*$
galaxies at $z=1$--$1.5$ \citep{Carilli2004}. 

\subsection{Comparing predictions by a theoretical model to inferred gas masses}
Semi-analytic models of galaxy formation start from a
  cosmological framework for structure formation, and include physical
  recipes to describe the baryonic physics acting on galaxies (such as the cooling and accretion of
  gas, star formation, and outflows). In this work we started from
  observed quantities and converted them into other quantities using
  empirical scaling laws. Although many of the ingredients are the
  same for the two approaches, their different nature makes them very
  complementary. Any tension between predictions from semi-analytic
  models and this work can shed light on the successes and failures of
  theoretical models.

We find that the \citet{Popping2013} semi-analytic model predicts three times less molecular
hydrogen than our inferred gas masses suggest at redshifts $0.5 < z <
1.0$. The semi-analytic model predicts  on average
about 2 times less molecular hydrogen at redshifts $1.0 < z <
2.5$. The same holds for galaxies at redshifts $2.5 < z < 3$ with stellar masses
less than $10^{10}\,\rm{M}_\odot$. Similarly, we find that the
semi-analytic model predicts less cold gas in galaxies at
redshifts $0.5 < z < 3$ than our inferred gas masses suggest.

Although the difference between semi-analytic model predictions and
our inferred gas masses is not huge (and well within the uncertainty range of
the method used in this work), it fits well within a broader picture where
theoretical models predict too little SF for galaxies at
redshifts $1 < z < 3$ \citep{Somerville2014}. If the molecular
hydrogen masses of modeled galaxies are too low for their stellar
mass, the SFRs of those galaxies will be too low as well when a
molecular gas based star-formation recipe is used. 

\citet{Popping2015} extended the sub-halo abundance matching approach
with recipes to infer cold gas masses. They found that the same semi-analytic
model as used in this work predicts less \emph{molecular hydrogen and cold gas}
at redshifts $1<z<3$ than their extended sub-halo abundance approach
suggests \citep[see also][]{Somerville2014, White2015}. A comparison
with different semi-analytic models
\citep{Obreschkow2009,Lagos2011cosmic_evol} yielded the same results,
making this a more general problem.

\subsection{What drives the star-formation efficiency of galaxies}
We have shown that, when using our model, the molecular fraction of
the cold gas is a very important quantity that cannot be neglected. Despite large total cold gas masses, low molecular
fractions can suppress the formation of stars and result in low
SFE.\footnote{We again emphasize that in this work the star-formation
  efficiency is defined as the ratio between SFR and total (\hi $+$
  \h2) cold gas mass, rather than just the molecular hydrogen mass} Indeed the SFEs of galaxies on the relation between SFR and
stellar mass increases with stellar mass \citep{Saintonge2013}. However, our model implies that the
physical state of the cold gas varies strongly as a
function of stellar mass along the stellar mass--SFR relation.

To understand the origin of the varying molecular fractions and SFE,
we have to look at the galaxy properties that enter our model. Within
our model the cold-gas and molecular-gas contents of a galaxy are set by a combination of its SFR, stellar mass, and size. Most important
is the combination of SFR and scale radius, which sets the SFR surface
density distribution of the galaxy. Within our model, low SFR surface
densities result in low \h2 surface densities. However, to ensure
the pressure of the gaseous disc is high enough for the cold gas to
collapse and form molecules, low \h2 surface densities
require relatively large cold-gas surface densities. This argument can
also be reversed: an increase in the scale radius of the gaseous disc lowers the surface
density of the cold gas, naturally causing less \h2 to self-shield and 
transform into stars. We see this in the galaxies with low stellar
masses and large discs in our sample. These
are galaxies with the highest gas fractions, but
their molecular fractions and star-formation efficiencies are very
low. 

An increase in galaxy scale radius also has a significant effect
on the SFE of galaxies with high stellar masses ($M_* >
10^{10}\,M_\odot$). At fixed stellar mass galaxies with compact discs
have higher SFEs (and molecular fractions) than galaxies with more
extended discs. The lower surface densities allow for a higher
shielding rate of the molecular hydrogen. At fixed stellar
mass the SFE of galaxies also increases with increasing SFR. The change in
SFE cannot account for differences in SFR of a few orders of
magnitude, but it can lower the SFR by a factor of a few. To first order it is the absolute amount of cold gas that
largely controls the gas density and SFR of a galaxy. The distribution
of gas into an extended disc can act as a secondary mechanism to suppress the formation of molecules and lower a galaxy's SFR.

In reality the formation of molecules does not
solely depend on the surface density or midplane pressure of cold gas. Dust grains
and metals are the primary catalysts and coolants of \h2
formation. Dust grains also shield the molecular hydrogen from \h2
dissociation by UV photons and photo-electric heating. The
metallicity of the ISM is therefore an important extra component
that controls the formation of molecules, especially in low-density
environments. In addition to low densities, a low-metallicity environment therefore stimulates
the build-up of large atomic gas reservoirs, preventing the cold gas
from condensing and forming stars \citep{KrumholzDekel2012,Berry2013,Popping2013,Somerville2015}. We have not included this effect in our model, but
it would strengthen the role that low surface densities and the partitioning of cold gas into
\hi and \h2 plays in the formation of stars. Furthermore, we have not
discussed the destruction of molecular hydrogen by radiation from an AGN.

\section{Summary \& Conclusions}
\label{sec:conclusion}
We applied a method to infer the total cold gas and
molecular gas content of a deep sample of galaxies from the
CANDELS survey covering a redshift range of $0.5 < z < 3$. All data
results in this paper are available for download
online.\footnote{\url{http://www.eso.org/~gpopping/Gergo_Poppings_Homepage/Data.html}}
Our main results are as follows:

\begin{itemize}
\item There is an increasing trend between the inferred cold gas, \hi, and \h2 mass and
  the stellar masses of galaxies, which is already in place at
  $z=3.0$ and extends down to stellar masses of $10^8\,\rm{M}_\odot$. The slopes of these trends are different from the slope
  of the relation between SFR and stellar mass.

\item There is no simple one-to-one mapping between inferred cold gas mass and
  molecular-hydrogen mass. Large gas reservoirs do not necessarily
  lead to large \h2 reservoirs. The molecular fractions of cold gas
  increase with increasing stellar mass and look-back time.

\item The cold gas fraction ($\rm{f}_{\rm gas}$) and relative amount
  of molecular hydrogen in galaxies (\frach2star) decrease gradually
  with time at a relatively constant rate, independent of stellar mass.

\item The mean fraction of atomic hydrogen ($M_{HI}/(M_* + M_{HI}$) in galaxies at fixed stellar mass stays remarkably constant
  with time over the entire redshift
  probed.

\item There is a large population of low-stellar-mass galaxies that
  are dominated by atomic gas. Only a minor fraction of their total
  gas content is molecular and can form stars.

\item The inferred SFE of galaxies increases along the relation between SFR and
  stellar mass. Although part of the same trend, the most-massive
  galaxies in our sample have SFEs more than twice as high as
  lower-mass galaxies. At fixed stellar mass the SFE and molecular
  fraction of the cold gas increase with galaxy compactness.

\item The adopted approach can be tremendously helpful in
  understanding the impact of selection biases for much smaller
  samples of directly observed gas masses, as well as extending
  scaling relation between gas mass and other galaxy properties to
  ranges and redshift difficult to reach through direct observations.

\end{itemize}

\section*{Acknowledgments}
We thank the referee for a thorough report and very
constructive comments that have improved the paper. GP thanks Romeel Dav\'e, Marco Spaans, and the galaxy group at the
Kapteyn Institute in Groningen for stimulating input for this
paper. This work is based on observations taken
by the CANDELS Multi-Cycle Treasury Program with
the NASA/ESA HST, which is operated by the Association of Universities for Research in Astronomy, Inc.,
under NASA contract NAS5-26555. GP
acknowledges NOVA (Nederlandse Onderzoekschool voor Astronomie) for funding.

\bibliographystyle{mn2e_fix}
\bibliography{references}

\end{document}